\documentclass[aps,twocolumn,amsmath,nofootinbib,longbibliography]{revtex4}
\usepackage{slashbox}
\usepackage{amsmath,amssymb,bm,mathrsfs,graphicx, braket, times,amsthm,enumerate, scrextend}
\usepackage[colorlinks=true,citecolor=blue,linkcolor=blue]{hyperref}
\usepackage{longtable}
\usepackage{multirow}
\usepackage{array}
\usepackage{bigstrut}
\usepackage[all,cmtip]{xy}
\usepackage[normalem]{ulem}
\usepackage[usenames,dvipsnames]{color}
\usepackage{makecell}
\usepackage{xcolor}
\newcommand{\noi}[1]{\noindent (#1)}
\begin{document}
\title{Exhaustive constructions of effective models in 1651 magnetic space groups}
\author{Feng Tang}\email{fengtang@nju.edu.cn}
\author{Xiangang Wan}\email{xgwan@nju.edu.cn}
\affiliation{National Laboratory of Solid State Microstructures and School of Physics, Nanjing University, Nanjing 210093, China and Collaborative Innovation Center of Advanced Microstructures, Nanjing University, Nanjing 210093, China}

\date{\today}

\maketitle

\textbf{
Even though the $k\cdot p$ effective Hamiltonian has been widely used to predict a large variety of phenomena in condensed matter systems, currently the common way to construct a $k\cdot p$ Hamiltonian is in a case-by-case and tedious manner. Moreover, deriving a $k\cdot p$ model in magnetic materials usually requires representation matrices  of a magnetic space group (MSG), however, those for type III/IV MSGs  are not well-developed compared with 230 type-I/II MSGs.  In this work, we explicitly tabulate all the representation matrices for all single-valued/double-valued irreducible representations (irreps) and co-irreps for the little groups of all special $k$ points in 1651 MSGs. Then through group theory analysis, we obtain a large database composed of 4 857 832 elementary $k\cdot p$ matrix blocks.  Directly using these matrix blocks, one can obtain any $k\cdot p$ Hamiltonian for any periodic system, for both nonmagnetic and magnetic materials.  In the light of  the huge successes achieved by $k\cdot p$ models, we believe that our work could bring about numerous advancements in various fields, such as semiconductors, topological physics, spintronics, valleytronics, twistronics etc. The exhaustive effective models, each attached to an MSG, would facilitate identification and realization of all possible emergent massless/massive excitations. More importantly, through comparing the theoretical model with novel property and our database, one can find the MSGs of materials realizations of the model.   We also expect our exhaustive results on $k\cdot p$ models to be utilized to connect other fields and condensed matter physics.}\newline

In many cases, the low-energy physical properties of periodic materials can be nicely described by the local dynamics around one or several $k$ point(s) in the Brillouin zone (BZ) \cite{kpbook,kpbook-2,TI-review-2010,Semimetal-review,HTC-review, berryphase-rmp}. Hence, a proper $k\cdot p$ Hamiltonian,  plays an important role in predictions of many intriguing properties, such as the electronic band structures of  semiconductors and their responses to external fields \cite{kpbook,kpbook-2}, the unconventional topologically protected surface and bulk states  in topological materials and their exotic transport behaviors \cite{Z2-TI-1,Z2-TI-2,Weyl,new-fermion,TI-review-2010,Semimetal-review, Bansil, Ando, valley}, etc. While the range of its validity is usually of an order of 10{\%} of the first BZ, the $k\cdot p$ model can be extended to describe the full-zone electronic band structure  by including a relatively large number of bands and high-order expansion \cite{kpbook}. Hence, an appropriate $k\cdot p$ Hamiltonian of large dimension could be a good starting point to derive a minimal model through various techniques \cite{HTC-review,OK-Anderen}.

Conventionally, $k\cdot p$ Hamiltonian was constructed based on perturbation theory using the Bloch eigenstates at $\mathbf{k}^*$ as the zeroth wavefunctions. Here $\mathbf{k}^*$ represents the $k$ vector in the BZ around which the $k\cdot p$ Hamiltonian is constructed. This scheme has been applied to derive  many well-known models in semiconductors, such as  Dresselhaus-Kip-Kittel model,  Kane model and Cardona-Pollak model for germanium and silicon crystals \cite{kpbook}, etc.   However, in deriving these models, the zeroth wavefunctions at $\mathbf{k}^*$ are usually originated from specific wavefunctions, e.g. atomic orbitals or plane wave functions, which may be more symmetric or biased, resulting in $k\cdot p$ Hamiltonians which  may miss nontrivial terms,  as we will demonstrate later. On the contrary, a $k\cdot p$ Hamiltonian obtained from a pure symmetry analysis through the method of invariants \cite{kpbook} naturally contain all possible symmetry-allowed terms without assuming concrete Bloch states, but purely derived from the symmetry properties of the participating Bloch states.  Practically,  one could prefer incorporating more bands to construct a symmetry-allowed $k\cdot p$ model which enlarges the range of validity.  Moreover, in many cases,   large-order $k\cdot p$ terms are required for a satisfactory description, e.g. the hexagonal warping effect in the surface states of topological insulator Bi$_2$Te$_3$ can be nicely interpreted by the $k\cdot p$ model up to at least third order of $\mathbf{q}$ ($\mathbf{q}\equiv \mathbf{k}-\mathbf{k}^*$) \cite{Bi2Te3-warp-PRL,Bi2Se3-warp-nature,Bi2Te3-warp-science}.   Unfortunately, the complexity of deriving a symmetry-allowed $k\cdot p$ Hamiltonian increases abruptly with increasing number of incorporated bands, the expansion order of $\mathbf{q}$ as well as the number of point symmetry operations in the little group $G(\mathbf{k}^*)$.  Hence, developing a database by which any $k\cdot p$ model can be quickly constructed  is of great significance and will definitely accelerate related studies  in condensed matter physics.

Though for 230 space groups, the (unitary) representation matrices of little groups, which are key to derive a symmetry-allowed $k\cdot p$ Hamiltonian,  have been listed explicitly \cite{bradley,bilbao} and have already attracted intensive interest as exemplified by the explosive studies of topological materials \cite{N-1,N-2,N-3}, those for other 674 type-III and 517 type-IV magnetic space groups (MSGs) have not. Besides,  even for 230 space groups, the representation matrix of antiunitary symmetry that could exist in the little group is not listed explicitly, though they should impose essential constraints on the theoretical model as shown in the following.

Recently, the 1651 MSGs, including  230 type-I,II space groups \cite{bradley} are getting more and more research interest  due to their promising applications  \cite{tokura, N-4}.  It is worth mentioning that  explicitly listing all little groups and their (co-)irreducible representations (irreps) for all special $k$ points in the 1651 MSGs necessarily help in deriving symmetry-allowed $k\cdot p$ models as well as construction of effective models in real space, namely, tight-binding models (or spin models). Besides,  it also benefits the searches for topological  magnetic materials \cite{tokura, N-4} and study of many other properties in magnetic materials, such as phase transitions in multiferroics, geometric phases, selection rules, degeneracies etc.  By providing a generic and unified formalism, symmetry analysis could reveal new  phenomena \cite{D} for which further derivations regarding specific mechanism can be inspired \cite{M}.

In this work, we firstly obtain all representation matrices for all single-valued/double-valued  (co-)irreps of little groups for all  special $k$ points in 1651 MSGs, and these data is shown in the Supplementary Material (SM) \cite{SM}: Part I.  By searching any MSG name in the Belov-Neronova-Smirnova (BNS) \cite{bns} notation in the index part of SM: Part I \cite{SM} and  clicking on the page number  corresponding to any special $k$ point of the MSG, one can find concrete operations of the little groups and the (co-)irrep matrices as well in all 1651 MSGs. Besides, how $k$ points in a $k$ star are related are also given in SM: Part I \cite{SM}.

To construct a $k\cdot p$ model, one first need to label the energy levels (denoted by $E_n$ or $E_{n'}$)  at $\mathbf{k}^*$ by (co-)irreps of the little groups, which can be calculated using the information (operations of little groups and characters of irreps)  as listed in SM: Part I \cite{SM} and wavefunctions by  first-principles calculations for realistic materials. The $k\cdot p$ Hamiltonian block $\mathcal{H}_{nn'}(\mathbf{q})$ is subject to constraints from (co-)irreps of $E_n$ and $E_{n'}$, as shown in Eq. \ref{hhh} in the following.    We then consider all possible doublets of (co-)irreps and  obtain the corresponding elementary $k\cdot p$ Hamiltonian blocks, which are sufficient to construct any $k\cdot p$ model, with any number of bands and any expansion order of $\mathbf{q}$ in principle.  In total, there are 4 857 832 such elementary $k\cdot p$ blocks  when choosing the cutoff of expansion order to be 4, which are all listed explicitly in SM: Part III \cite{SM}. Our exhaustive results  can be applied to metals and insulators (semiconductors). All possible excitations, massless or massive,   can be studied using the constructed $k\cdot p$ models, like Dirac equations in topological insulators \cite{Z2-TI-1,Z2-TI-2} and those for  ``Dirac octets''  in antiperovskites \cite{octet}.  We tabulate all linear $k\cdot p$ models around all band crossings in SM: Part II which could facilitate the studies in topological semimetals \cite{Semimetal-review}.   The exhaustive $k\cdot p$ blocks can also be used to solve the inverse problem of finding MSG(s) of materials realization corresponding to a targeted $k\cdot p$ Hamiltonian. Our $k\cdot p$ results attached to MSGs, make their materials realizations very efficient by directly screening  materials crystallizing in the corresponding MSG(s) in materials databases.

In the following, we show how to obtain and use the elementary $k\cdot p$ Hamiltonian blocks listed in SM: Part III \cite{SM} to construct any $k\cdot p$ Hamiltonian with expansion order denoted by $L$. Hereafter we omit the expansion order, but the reader should keep in mind that all formulas deal with a fixed expansion order $L=0,1,2,3,4,\ldots$   We distinguish between two cases: (case I) there is no antiunitary symmetry in the little group $G(\mathbf{k}^*)$ and (case II) there exists at least one antiunitary symmetry. In case I, we use the irreps of $G(\mathbf{k}^*)$ to label the energy levels at $\mathbf{k}^*$ and different irreps are denoted by different Greek letters $\alpha,\beta,\gamma,\delta$, which can take $1,2,3,\ldots$. The representation matrices $D^\alpha(h),D^\beta(h),D^\gamma(h),D^\delta(h) (h\in H(\mathbf{k}^*)$, the unitary part of little group, namely, $H(\mathbf{k}^*)=G(\mathbf{k}^*))$ for these irreps are listed in SM: Part I \cite{SM}. In case II, we should use co-irreps to label each energy level which are actually originated from the irreps of  the unitary subgroup of $G(\mathbf{k}^*)$. This unitary group is denoted as $H(\mathbf{k}^*)$ and $G(\mathbf{k}^*)=H(\mathbf{k}^*)+A\cdot H(\mathbf{k}^*)$. Here $A$ can be expressed by $\{\beta_a|\boldsymbol{\tau}_a\}\Theta$ where $\Theta$ is the time-reversal operator while  $\{\beta_a|\boldsymbol{\tau}_a\}$ is a spatial operation in Seitz notation.  The point part $\beta_a$ can always be chosen to be a two-order operation, such as 2-fold rotation $C_2$, mirror $m$ and inversion $I$. The same as case I, the irreps of $H(\mathbf{k}^*)$ are also represented by $\alpha,\beta,\ldots$, while the co-irreps of $G(\mathbf{k}^*)$ can be in the form of $\{\alpha\}$, $\{\beta,\beta\}$, or $\{\gamma,\delta\}$ (see Sec. \ref{app:Aonirreps} of the Appendix) corresponding to three types of consequences of $A$ on irreps of $H(\mathbf{k}^*)$ \cite{bradley}. In both cases, we use $\xi_n$ to represent the irrep or co-irrep of $E_n$.

Before we describe the constraints from unitary symmetries in $H(\mathbf{k}^*)$ in both cases I and II, we expand $\mathcal{H}_{nn'}(\mathbf{q})$  in the following form (see Sec. \ref{q-basis} of the Appendix for details):
\begin{equation}\label{W=f}
    \mathcal{H}_{nn'}(\mathbf{q})=\sum_{l}\mathcal{H}_{nn'}^{l}q^{L-l},
\end{equation}
where $l$ should take a set of non-negative integers $l=L,L-2,L-4,\ldots,$ ($l\ge0$), $q=|\mathbf{q}|$ and $\mathcal{H}^l_{nn'}$ can be expressed by polynomial basis functions shown in Table \ref{tab2} of the Appendix.

\noindent\textbf{Constructing elementary $k\cdot p$ Hamiltonian blocks in case I.}\\
 For case I, $\xi_n$ can be expressed as $\alpha_n$ representing  an irrep of $G(\mathbf{k}^*)=H(\mathbf{k}^*)$ for energy level $n$. The constraints on $\mathcal{H}^l_{nn'}$ (written as $\mathcal{H}^l_{\alpha_n,\alpha_{n'}}$) from $H(\mathbf{k}^*)$ are as follows:
\begin{equation}\label{hhh}
    D^{\alpha_n}(h)\mathcal{H}^l_{\alpha_n \alpha_{n'}}(h_0^{-1}\mathbf{q}){D^{\alpha_{n'}}(h)}^\dag=\mathcal{H}^l_{\alpha_{n}\alpha_{n'}}(\mathbf{q}), \forall h \in H(\mathbf{k}^*),
\end{equation}
where   $h_0$ denotes the point group operation in $h$.

From Eq. \ref{hhh}, it is clear that the form of $\mathcal{H}^l_{nn'}$ is uniquely determined by the irreps $\alpha_n$ and $\alpha_{n'}$.  The solution of $\mathcal{H}^{l}_{\alpha_n\alpha_{n'}}$ satisfying Eq. \ref{hhh} can be written as the following combination:
\begin{equation}\label{W=H}
    \mathcal{H}_{\alpha_n,\alpha_{n'}}^l(\mathbf{q})=\sum_{m}a_m H^l_{\alpha_n,\alpha_{n'},m}(\mathbf{q}),
\end{equation}
where $a_m$ is a nonzero coefficient and can be tuned freely. $\{H^l_{\alpha_n,\alpha_{n'},m}\}_{m=1,2,\ldots}$ are the elementary $k\cdot p$ blocks in case I and are explicitly given in SM: Part III \cite{SM} following the indicator $\alpha_n\&\alpha_{n'}$. Note that when Eq. \ref{hhh} has no nontrivial solution other than  $\mathcal{H}^{l}_{\alpha_n\alpha_{n'}}=0$,  $\{H^l_{\alpha_n,\alpha_{n'},m}\}_{m}=\{\}$  for which no results are given  in SM: Part III \cite{SM}.

In addition, when $\alpha_{n}=\alpha_{n'}$, we can further require that $H^l_{\alpha_n,\alpha_{n},m}(\mathbf{q})$ is Hermitian for any $m$. Note that the $k\cdot p$ block $\mathcal{H}_{nn'}$ also satisfies: If  $n=n'$, namely $H_{nn'}$ is a diagonal block, it should be Hermitian, and  the coefficients in Eq. \ref{W=H} should be real, namely $a_m\in\mathbb{R}$.  If $n\ne n'$ (nondiagonal block), $\mathcal{H}_{nn'}=\mathcal{H}_{n'n'}^\dag$, and  the coefficients in Eq. \ref{W=H} are complex, namely $a_m\in\mathbb{C}$.

Here, with no loss of generality, we show an example of MSG 195.1 (type I), and choose three energy levels $E_a,E_b,E_c$ at $\Gamma$ point of which $E_a$ corresponds to the first single-valued irrep while the rest two correspond to the fourth single-valued irrep. When $L=1$,  $l$ thus can only take $1$ according to Eq. \ref{W=f}.  As shown in Eq. \ref{W=H},  $\mathcal{H}^{l=1}_{bc}$ is the linear combination of the elementary $k\cdot p$ blocks $\{H^{l=1}_{4,4,m}\}$. By reference to SM: Part III \cite{SM}, for $\Gamma$ point in MSG 195.1, $\{H^{l=1}_{4,4,m}\}$ (shown after indicator $4\&4$ in SM: Part III \cite{SM}) contains two elementary $k\cdot p$ blocks (thus $m=1,2$): $H^{l=1}_{4,4,1}=P_3 S_{\{1,2\}}+P_2 S_{\{1,3\}}+P_1 S_{\{2,3\}}=\frac{1}{\sqrt 2}\left(\begin{array}{ccc}0& q_z&  q_y\\ q_z& 0& q_x\\ q_y&q_x& 0\end{array}\right)$ and $H^{l=1}_{4,4,2}=-P_1 A_{\{2,3\}}+P_2 A_{\{1,3\}}-P_3 A_{\{1,2\}}=\frac{1}{\sqrt 2}\left(\begin{array}{ccc}0& \mathrm{i}q_z&  -\mathrm{i}q_y\\ -\mathrm{i}q_z& 0& \mathrm{i}q_x\\ \mathrm{i}q_y&-\mathrm{i}q_x& 0\end{array}\right)$  where $\mathrm{i}^2=-1$. See the definition of $P_i$ and $A_{\{i,j\}}$, $S_{\{i,j\}}$  in Secs. \ref{q-basis} and \ref{m-basis} of the Appendix, respectively, and we also show these $3\times 3$  matrices $A_{\{i,j\}}$ and $S_{\{i,j\}}$  explicitly in the caption of Table \ref{tab:36}. According to Eq. \ref{W=H}, $\mathcal{H}^{l=1}_{bc}=c_1 H^{l=1}_{4,4,1}+c_2 H^{l=1}_{4,4,2}=\frac{1}{\sqrt 2}\left(\begin{array}{ccc}0& c_+q_z&  c_-q_y\\ c_-q_z& 0& c_+q_x\\ c_+q_y&c_-q_x& 0\end{array}\right)$ where $c_{\pm}=c_1\pm\mathrm{i}c_2$ and  $c_1$ and $c_2$ are two complex parameters.   For $\mathcal{H}^l_{bb}$, a diagonal block, it can also be constructed using  $H^{l=1}_{4,4,1}$ and $H^{l=1}_{4,4,2}$, but with two real parameters $r_1$ and $r_2$: $\mathcal{H}^{l=1}_{bb}=\frac{1}{\sqrt 2}\left(\begin{array}{ccc}0& r_+q_z&  r_-q_y\\ r_-q_z& 0& r_+q_x\\ r_+q_y&r_-q_x& 0\end{array}\right)$ where $r_{\pm}=r_1\pm\mathrm{i}r_2$,  which could realize spin-1 fermionic excitations \cite{new-fermion}.  The rest $k\cdot p$ blocks like $\mathcal{H}^{l=1}_{aa}$, $\mathcal{H}^{l=1}_{ab}, \ldots$  can be found in Sec. \ref{app:195} of the Appendix where we also derive the whole $k\cdot p$ Hamiltonian up to second order (namely, including contributions from $k\cdot p$ Hamiltonians when $L=0,1,2$).

\noindent\textbf{Elementary $k\cdot p$ Hamiltonian blocks in case II}\\
 In case II, other than the constraints from $H(\mathbf{k}^*)$ as shown in Eq. \ref{hhh}, $\mathcal{H}_{nn'}$ should be subject to  additional constraint from $A$. The details of deriving elementary $k\cdot p$ blocks are shown in Sec. \ref{app:kpAppendix} of the Appendix. We suggest the readers who are not concerned with these details  can directly consult Tables \ref{A} and \ref{AA} in the Appendix, which are enough to construct any $k\cdot p$ model in case II.

Table \ref{A} in the Appendix shows all possible doublets of $\xi_n\&\xi_{n'}$. If $\xi_n=\xi_{n'}$, there would be three possibilities corresponding to $\{\alpha\},\{\beta,\beta\}$ and $\{\gamma,\delta\}$.  For $\xi_n=\xi_{n'}=\{\alpha\}$, as shown in Table \ref{A}, there are two types of elementary $k\cdot p$ blocks, $\{\bar{H}^l_{\alpha\alpha\bar{m}}\}$ and $\{\bar{H'}^l_{\alpha\alpha\bar{m}'}\}$, corresponding to the cases that $n=n'$ and $n\ne n'$, respectively, since for them, the additional constraints by $A$ are different.  The corresponding explicit formulas of constructing $\mathcal{H}_{nn'}^l$ can be found in Table \ref{AA} in the Appendix, which clearly shows that these two types of elementary $k\cdot p$ blocks  are used in diagonal and nondiagonal blocks, respectively.

 When $\xi_n=\xi_{n'}=\{\beta,\beta\}$, there are also two types of elementary $k\cdot p$ blocks, $\{H^l_{\beta\beta m}\}$ and $\{\bar{H}^l_{\beta\beta \bar{m}}\}$, as shown in Table \ref{A} in the Appendix. When $n\ne n'$,   $\{H^l_{\beta\beta m}\}$, is enough to construct $\mathcal{H}^l_{nn'}$ as seen from Table \ref{AA} in the Appendix. When $n=n'$, both types of elementary $k\cdot p$ blocks should be used.

 When $\xi_n=\xi_{n'}=\{\gamma,\delta\}$, there are three types of elementary $k\cdot p$ blocks,  $\{H^l_{\gamma\gamma m}\}$, $\{H^l_{\gamma\delta m}\}$ and $\{\bar{H}^l_{\gamma\delta\bar{m}}\}$. The construction of diagonal block (namely, $n=n'$) need all these types while that for the nondiagonal block (namely, $n\ne n'$) only need the former two types ($\{H^l_{\gamma\gamma m}\}$, $\{H^l_{\gamma\delta m}\}$) as shown in Table \ref{AA} in the Appendix.

Similarly,  when $\xi_n\ne\xi_{n'}$ for which $n$ is definitely not equal to $n'$,  all elementary $k\cdot p$ blocks needed are listed in Table \ref{A} in the Appendix and how they are used to construct $k\cdot p$ block $\mathcal{H}^l_{nn'}$ is also given in Table \ref{AA} in the Appendix.

 Note that all the above different types of elementary $k\cdot p$ blocks are given in SM: Part III \cite{SM} printed in the same colors as in Table \ref{A} in the Appendix. For case II, we take  the construction of $k\cdot p$ model around $\Gamma$ point of MSG 216.75 as the example  in the following.\\

\noindent \textbf{Example: $k\cdot p$ model in MSG 216.75.}\\
Here demonstrate an example of constructing first-order $k\cdot p$ Hamiltonian around $\Gamma$ point in zinceblende structures.    The space group number for the zincblende structure is 216 with time-reversal symmetry (TRS)   considered (the MSG is thus 216.75, of type II). As shown in SM: Part I \cite{SM}, the $\Gamma$ point in MSG 216.75 owns three  different double-valued co-irreps, denoted by $\{1\},\{2\}$ and $\{3\}$, whose dimensions are $2, 2$ and $4$, respectively.

We  consider three energy levels: $E_a=\epsilon_1,E_b=\epsilon_2,E_c=\epsilon_3$ at $\Gamma$ point to construct the $k\cdot p$ model. Here we use $a,b,c$ to denote the band index $n$ or $n'$ thus there are 9 $k\cdot p$ blocks which can be written as $\mathcal{H}_{aa}, \mathcal{H}_{ab}, \mathcal{H}_{ac}, \mathcal{H}_{ba}, \ldots$. The co-irreps of these energy levels are chosen to be: $\xi_a=\{1\}, \xi_b=\{2\}, \xi_c=\{3\}$. For the first-order $k\cdot p$ model, namely, $L=1$, thus $l$ can only take $1$ by Eq. \ref{W=f}. To construct the diagonal $k\cdot p$ blocks $\mathcal{H}_{aa}, \mathcal{H}_{bb}$ and $\mathcal{H}_{cc}$, we should use the elementary $k\cdot p$ blocks $\{\bar{H}^{l=1}_{11\bar{m}}\}$, $\{\bar{H}^{l=1}_{22\bar{m}}\}$, and $\{\bar{H}^{l=1}_{33\bar{m}}\}$ as shown in Table \ref{AA}. They are given in SM: Part III \cite{SM} printed in purple (as shown in Table \ref{A}), following the indicators $\{1\}\&\{1\}$, $\{2\}\&\{2\}$ and $\{3\}\&\{3\}$, respectively, corresponding to $l=1$ part. To construct the non-diagonal blocks:  $\mathcal{H}_{ab}, \mathcal{H}_{ac}$ and $\mathcal{H}_{bc}$, we should use the elementary $k\cdot p$ blocks as $\{\bar{H}^{l=1}_{12\bar{m}}\}$, $\{\bar{H}^{l=1}_{13\bar{m}}\}$, and $\{\bar{H}^{l=1}_{23\bar{m}}\}$, which are given in SM: Part III \cite{SM} printed in black,  following the indicators $\{1\}\&\{2\}$, $\{1\}\&\{3\}$ and $\{2\}\&\{3\}$, respectively.  Combining these $k\cdot p$ blocks,  we could obtain the $k\cdot p$ Hamiltonian to the first order as shown in Eq. \ref{examples-2} of the Appendix, with more details  left to Sec. \ref{app:detail-216} of the Appendix.

After making unitary transformation for the sake of comparison with the first-order eight-band Kane model \cite{kpbook}, which deals with the same structure but using special basis sets at $\Gamma$ (see Eq. \ref{app:kane-basis} of the Appendix), we obtain our $k\cdot p$ Hamiltonian  as shown in  Eq. \ref{H216} of the following.
\begin{widetext}
\begin{equation}\label{H216}
\mathcal{H}(\mathbf{q})=\left(
\begin{array}{cccccccc}
 \epsilon_2 & r_4 q_z & -\frac{1}{2} \sqrt{3} r_4 q_+ & -r_2 q_z & 0 & \frac{r_4 q_-}{2 } & 0 &  r_2 q_- \\
 r_4 q_z & \epsilon_3 & - r_1 q_- & 0 & -\frac{r_4 q_-}{2 } &  \sqrt{3} r_1 q_+ &  2 r_1 q_z & \frac{1}{2} \sqrt{3} r_3 q_+ \\
 -\frac{1}{2} \sqrt{3} r_4 q_- & -r_1 q_+ & \epsilon_3 & \frac{r_3 q_+}{2} & 0 & 2 r_1 q_z & \sqrt{3} r_1 q_- & -r_3 q_z \\
 -r_2 q_z & 0 & \frac{r_3 q_-}{2} & \epsilon_1 & -r_2 q_- & \frac{1}{2} \sqrt{3} r_3 q_+ & -r_3 q_z & 0 \\
 0 & -\frac{r_4 q_+}{2 } & 0 & -r_2 q_+ & \epsilon_2 & r_4 q_z & -\frac{1}{2} \sqrt{3} r_4 q_- & -r_2 q_z \\
 \frac{r_4 q_+}{2} &  \sqrt{3} r_1 q_- & 2r_1 q_z & \frac{1}{2} \sqrt{3} r_3 q_- &  r_4 q_z & \epsilon_3 &  r_1 q_+ & 0 \\
 0 & 2r_1 q_z & \sqrt{3} r_1 q_+ & -r_3 q_z & -\frac{1}{2} \sqrt{3} r_4 q_+ & r_1 q_- & \epsilon_3 & -\frac{r_3 q_-}{2} \\
 r_2 q_+ & \frac{1}{2} \sqrt{3} r_3 q_- & -r_3 q_z & 0 & -r_2 q_z & 0 & -\frac{r_3 q_+}{2} & \epsilon_1 \\
\end{array}
\right),\end{equation}
\end{widetext}
where $q_{\pm}=q_x\pm \mathrm{i} q_y$ and $r_1$, $r_2$, $r_3$ and $r_4$ are all real parameters.

The  first order eight-band Kane Hamiltonian $H_{Kane}$ (shown in Eq. \ref{app:kane} of the Appendix)  considers three energy levels around the Fermi level at $\Gamma$ points originated from $s$ and $p$ atomic orbitals  and the concrete expression of corresponding  eigenstates is shown in  Sec. Eq. \ref{app:kane-basis} of the Appdendix. According to these eigenstates, the co-irreps of these energy levels are found to be  $\{1\},\{2\},\{3\}$.  It is straightforward to find that, when $\epsilon_3\rightarrow 0, \epsilon_2\rightarrow E_0, \epsilon_1\rightarrow-\Delta_0, r_4\rightarrow -\sqrt{\frac{2}{3}}P, r_2\rightarrow-\frac{1}{\sqrt 3}P$ and $r_1,r_3\rightarrow0$,  $\mathcal{H}(\mathbf{q})$ in Eq. \ref{H216} would change to the first-order Kane model ($H_{Kane}$). Thus $\mathcal{H}$ in Eq. \ref{H216} would own 3 more real parameters.  The reason that $r_1,r_3$ are vanishing and $r_4=\sqrt 2 r_2$ in $H_{Kane}$ can be easily understood  since the atomic orbitals are used in deriving $H_{Kane}$ which are over-symmetrized obviously since space group 216 has no inversion center. In  many semiconductors of III-V family \cite{kane-8}, $r_1$ and $r_3$   may be very small and $r_4\sim \sqrt2 r_2$ so that $H_{Kane}$ can still be a good approximation. However, since $r_1$ and $r_3$ as well as the independence of $r_2$ and $r_4$ are allowed by symmetry arguments, it is interesting to reveal their effects  in future studies.

Other than the above example, in Sec. \ref{gra} of the Appendix, we also take graphene (MSG 191. 234)  as an example to show how to construct $k\cdot p$ models in case II combining Tables \ref{A} and \ref{AA} of the Appendix and using the corresponding elementary $k\cdot p$ blocks in SM: Part III \cite{SM}.

Besides,  by considering all different co-irreps  in the form of $\{\beta,\beta\}$ of high-symmetry line $Q$ of MSG 62.442,  we construct an eight-band $k\cdot p$ model to second order shown in Eq. \ref{62}  of the Appendix.   To our best knowledge, this is the first model constructed from concrete MSG  which can realize a Hopf-link nodal structures  \cite{hopf-link-1,hopf-link-2,hopf-link-3} and this model could be applied in subsequent theoretical studies of electronic properties, including transport and optical behaviors under external fields of electronic states around the Hopf-link nodal loops, surface states and the interaction induced instability etc.

\begin{widetext}
\begin{table*}[!tbhp]
\begin{tabular}{c|c|c|c|c|c}
\hline\hline
  Type& MSG & $\mathbf{k}^*$ name & $\mathbf{k}^*$ coordinate& degeneracy  & linear $k\cdot p$ model\\\hline
  III& 207.42& $\Gamma$& $(0,0,0)$ & 3& $r_1(A_{\{1,2\}}q_z-A_{\{1,3\}}q_y+A_{\{2,3\}}q_x)+r_2(S_{\{1,2\}}q_z+S_{\{1,3\}}q_y+S_{\{2,3\}}q_x)$ \\\hline
  III& 207.42& $R$& $(\frac{1}{2},\frac{1}{2},\frac{1}{2})$ & 3& $r_1(A_{\{1,2\}}q_z+A_{\{1,3\}}q_y+A_{\{2,3\}}q_x)+r_2(S_{\{1,2\}}q_z-S_{\{1,3\}}q_y+S_{\{2,3\}}q_x)$\\\hline
  III& 208.46& $\Gamma$& $(0,0,0)$ & 3& $r_1(A_{\{1,2\}}q_z+A_{\{1,3\}}q_y+A_{\{2,3\}}q_x)+r_2(S_{\{1,2\}}q_z-S_{\{1,3\}}q_y+S_{\{2,3\}}q_x)$\\\hline
    III& 208.46& $R$& $(\frac{1}{2},\frac{1}{2},\frac{1}{2})$ & 3& $r_1(A_{\{1,2\}}q_z-A_{\{1,3\}}q_y-A_{\{2,3\}}q_x)+r_2(S_{\{1,2\}}q_z-S_{\{1,3\}}q_y-S_{\{2,3\}}q_x)$\\\hline
      III& 209.50& $\Gamma$& $(0,0,0)$ & 3& $r_1(A_{\{1,2\}}q_z+A_{\{1,3\}}q_y+A_{\{2,3\}}q_x)+r_2(S_{\{1,2\}}q_z-S_{\{1,3\}}q_y+S_{\{2,3\}}q_x)$\\\hline
      III& 210.54& $\Gamma$& $(0,0,0)$ & 3& $r_1(A_{\{1,2\}}q_z-A_{\{1,3\}}q_y+A_{\{2,3\}}q_x)+r_2(S_{\{1,2\}}q_z+S_{\{1,3\}}q_y+S_{\{2,3\}}q_x)$\\\hline
         III& 211.58& $\Gamma$& $(0,0,0)$ & 3& $r_1(A_{\{1,2\}}q_z-A_{\{1,3\}}q_y-A_{\{2,3\}}q_x)+r_2(S_{\{1,2\}}q_z+S_{\{1,3\}}q_y-S_{\{2,3\}}q_x)$\\\hline
          III& 211.58& $H$& $(\frac{1}{2},-\frac{1}{2},\frac{1}{2})$ & 3& $r_1(A_{\{1,2\}}q_z-A_{\{1,3\}}q_y+A_{\{2,3\}}q_x)+r_2(S_{\{1,2\}}q_z+S_{\{1,3\}}q_y+S_{\{2,3\}}q_x)$\\\hline
           III& 211.58& $P$& $(\frac{1}{4},\frac{1}{4},\frac{1}{4})$ & 3& $r_1(A_{\{1,2\}}q_z-A_{\{1,3\}}q_y+A_{\{2,3\}}q_x)+r_2(S_{\{1,2\}}q_z+S_{\{1,3\}}q_y+S_{\{2,3\}}q_x)$\\\hline
            III& 211.58& $P$& $(-\frac{1}{4},-\frac{1}{4},\frac{3}{4})$ & 3& $r_1(A_{\{1,2\}}q_z+A_{\{1,3\}}q_y-A_{\{2,3\}}q_x)+r_2(S_{\{1,2\}}q_z-S_{\{1,3\}}q_y-S_{\{2,3\}}q_x)$\\\hline
          III& 212.61& $\Gamma$& $(0,0,0)$ & 3& $r_1(A_{\{1,2\}}q_z+A_{\{1,3\}}q_y-A_{\{2,3\}}q_x)+r_2(S_{\{1,2\}}q_z-S_{\{1,3\}}q_y-S_{\{2,3\}}q_x)$\\\hline
         III& 213.65& $\Gamma$& $(0,0,0)$ & 3& $r_1(A_{\{1,2\}}q_z-A_{\{1,3\}}q_y+A_{\{2,3\}}q_x)+r_2(S_{\{1,2\}}q_z+S_{\{1,3\}}q_y+S_{\{2,3\}}q_x)$\\\hline
          III& 214.69& $\Gamma$& $(0,0,0)$ & 3& $r_1(A_{\{1,2\}}q_z+A_{\{1,3\}}q_y-A_{\{2,3\}}q_x)+r_2(S_{\{1,2\}}q_z-S_{\{1,3\}}q_y-S_{\{2,3\}}q_x)$\\\hline
              III& 214.69& $H$& $(\frac{1}{2},-\frac{1}{2},\frac{1}{2})$ & 3& $r_1(A_{\{1,2\}}q_z-A_{\{1,3\}}q_y-A_{\{2,3\}}q_x)+r_2(S_{\{1,2\}}q_z+S_{\{1,3\}}q_y-S_{\{2,3\}}q_x)$\\\hline
              III& 217.80& $P$& $(\frac{1}{4},\frac{1}{4},\frac{1}{4})$ & 3& $r_1(A_{\{1,2\}}q_z-A_{\{1,3\}}q_y+A_{\{2,3\}}q_x)+r_2(S_{\{1,2\}}q_z+S_{\{1,3\}}q_y+S_{\{2,3\}}q_x)$\\\hline
              III& 217.80& $P$& $(-\frac{1}{4},-\frac{1}{4},\frac{3}{4})$ & 3& $r_1(A_{\{1,2\}}q_z-A_{\{1,3\}}q_y-A_{\{2,3\}}q_x)+r_2(S_{\{1,2\}}q_z+S_{\{1,3\}}q_y-S_{\{2,3\}}q_x)$\\\hline
              III& 229.143& $P$& $(\frac{1}{4},\frac{1}{4},\frac{1}{4})$ & 3& $r_1(A_{\{1,2\}}q_z-A_{\{1,3\}}q_y+A_{\{2,3\}}q_x)+r_2(S_{\{1,2\}}q_z+S_{\{1,3\}}q_y+S_{\{2,3\}}q_x)$\\\hline
                 IV& \makecell[c]{195.3,
                  196.6,\\
                  198.11,
                  207.43, \\
                208.47,
                209.51, \\
               210.55,
                212.62, \\
                213.66}& $\Gamma$& $(0,0,0)$ & 3& $r_1(A_{\{1,2\}}q_z-A_{\{1,3\}}q_y+A_{\{2,3\}}q_x)$\\\hline
                IV& 195.3& $R$& $(1/2,1/2,1/2)$ & 6&  $\left(\begin{array}{cc}h^{207.42}& c_1(S_{\{1,2\}}q_z+S_{\{1,3\}}q_y+S_{\{2,3\}}q_x)\\h.c.& -{h^{207.42}}^* \end{array}\right)$\\\hline
                 IV& 207.43,208.47& $R$& $(\frac{1}{2},\frac{1}{2},\frac{1}{2})$ & 6&  $\left(\begin{array}{cc}h^{195.3}& c_1(S_{\{1,2\}}q_z+A_{\{1,3\}}q_y+S_{\{2,3\}}q_x+\mathrm{dia}(0,\sqrt 2q_z,\sqrt 2 q_y))\\h.c.&  h^{195.3} \end{array}\right)$\\\hline\hline

\end{tabular}
\caption{The list of concrete $k$ points realizing chiral three-fold or six-fold degenerate fermions in type III and IV MSGs. The four columns contain type of MSG, name of $k$ point, coordinate of $k$ point and the concrete linear model, respectively. The MSGs are given in the BNS notation.  The SOC is negligible.  The $3\times 3$ matrices: $A_{\{i,j\}}$ or $S_{\{i,j\}}$ ($i<j$) are all Hermitian and normalized, and are antisymmetric or symmetric, respectively, as described in Eq. \ref{m-basis} of the Appendix: The subscript $\{i,j\}$ denotes the nonzero entries $(i,j)$ and $(j,i)$ while all the rest entries are vanishing. Concretely, $A_{\{1,2\}}=\frac{1}{\sqrt 2}\left(\begin{array}{ccc}0& -\mathrm{i}& 0\\ \mathrm{i}& 0& 0\\0& 0& 0\end{array}\right)$, $A_{\{1,3\}}=\frac{1}{\sqrt 2}\left(\begin{array}{ccc}0& 0& -\mathrm{i}\\ 0& 0& 0\\ \mathrm{i}& 0& 0\end{array}\right)$ and $A_{\{2,3\}}=\frac{1}{\sqrt 2}\left(\begin{array}{ccc}0& 0& 0\\ 0& 0& -\mathrm{i}\\0& \mathrm{i}& 0\end{array}\right)$. $S_{\{1,2\}}=\frac{1}{\sqrt 2}\left(\begin{array}{ccc}0& 1& 0\\ 1& 0& 0\\0& 0& 0\end{array}\right)$, $S_{\{1,3\}}=\frac{1}{\sqrt 2}\left(\begin{array}{ccc}0& 0& 1\\ 0& 0& 0\\1& 0& 0\end{array}\right)$ and $S_{\{2,3\}}=\frac{1}{\sqrt 2}\left(\begin{array}{ccc}0& 0& 0\\ 0& 0& 1\\0& 1& 0\end{array}\right)$.   $h^{195.3}$ denotes the linear model for $\Gamma$ point in MSG 195.3, namely $h^{195.3}=r_1(A_{\{1,2\}}q_z-A_{\{1,3\}}q_y+A_{\{2,3\}}q_x)$ and $h^{207.42}$ denotes the linear model for $\Gamma$ point in MSG 207.42: $h^{207.42}=r_1(A_{\{1,2\}}q_z-A_{\{1,3\}}q_y+A_{\{2,3\}}q_x)+r_2(S_{\{1,2\}}q_z+S_{\{1,3\}}q_y+S_{\{2,3\}}q_x)$.}
\label{tab:36}
\end{table*}
\end{widetext}

\noindent \textbf{$k\cdot p$ models of all band crossings.}\\
The $k\cdot p$ models around various band crossings play important roles  in the study of the topological character and intriguing properties in topological semimetals \cite{Weyl, new-fermion,KramersWeyl,Manes,SMYoung,DoubleDSM}, however, the previous studies were almost all based on 230 space groups. On the other hand, the studies of topological band crossings in magnetically-ordered systems are attracting extensive interest due to their promising applications in spintronics  with existing nontrivial topology and magnetism \cite{tokura, N-4}. While multifold fermions around high-symmetry points in type-III/IV MSGs  considering  spin-orbit coupling (SOC) have been discussed \cite{APL}, tabulating all possible $k\cdot p$ models of band crossings  in all MSGs and both symmetry settings (considering SOC or not) would be very significant, which could demonstrate accurate topological behaviors of excitations and also provide a useful guide for realizing various topological semimetals \cite{Semimetal-review}.

Using SM: Part I and Part III \cite{SM},  we  could thus  obtain a complete list of all $k\cdot p$ models (to the 4th order of $\mathbf{q}$ at most) of  band crossings.  For high-symmetry points, the band crossing corresponds to one degenerate (co-)irrep, while for high-symmetry lines, two different (co-)irreps should be considered to constitute a band crossing.   We explicitly list those linear $k\cdot p$ models in SM: Part II \cite{SM}. To demonstrate,   we list the $k\cdot p$ models for those chiral  3-fold/6-fold degenerate band crossings at the  high-symmetry points in types-III/IV MSGs with SOC negligible in Table \ref{tab:36}, which are  rarely studied compared with 230 types-I/II MSGs. Here we firstly filter out high-symmetry points whose point group symmetry contains improper operation to ensure chirality \cite{KramersWeyl}, which can be easily done by checking the little groups shown in SM: Part I \cite{SM}. It is worth mentioning that the MSGs listed in Table \ref{tab:36} could also be utilized in bosonic systems, such as magnons \cite{cuteo}.

With various degeneracies ($\le 8$), the band crossings   include  Weyl points \cite{Weyl}, Dirac points \cite{Na3Bi, Cd3As2, SMYoung}, triply-degenerate points and others \cite{new-fermion}, etc. Since  our expansion of $k\cdot p$ model can be up to the fourth order of $\mathbf{q}$, band crossings with high topological charges \cite{Fang-highChern} can also be identified directly. These excitations, whose MSGs are clearly known, are easy to realize in crystalline materials combined with first-principles calculations. The explicit $k\cdot p$ models for band crossings in realistic materials are of important applied value since they are expected  to own large Berry curvature and anomalous transport behaviors  \cite{berryphase-rmp} and act as basis to fabricate layered structures realizing novel phenomena like nonlinear Hall effect \cite{nonlinear-Hall}.

\noindent \textbf{Realization of a prior model: real-Dirac fermion.}\\
Our elementary $k\cdot p$ blocks could also be applied to solve the inverse problem of identifying MSG to which materials belong could realize an interesting theoretical model that was proposed previously but lacks (good) materials realizations. Such identification is simply through straightly comparing with the numerous and exhaustive  elementary $k\cdot p$ blocks shown in SM: Part III \cite{SM}.  When a candidate MSG is found, the theoretical Hamiltonian can thus be realized in materials crystallizing in the MSG.  Besides, if it is found  that no MSGs could realize a targeted model, this model can thus only be realized through fine tuning parameters in crystalline materials or in cold-atom systems.

We present an example here: searching for realizations of real Dirac fermions,  which are the real generalizations of Weyl fermions. The real Dirac fermions were proposed theoretically \cite{real-dirac} but lack materials realizations.  Here we start from the Hamiltonian  required by the real Dirac fermions as follows:
\begin{equation}\label{rd-H}
    H_{RD}(\mathbf{q})=v_z q_z \sigma_z\otimes\tau_0+ v_x q_x \sigma_x\otimes\tau_x +v_y q_y \sigma_x\otimes\tau_z,
\end{equation}
where $\sigma_i$ and $\tau_i$ are Pauli matrices and $v_x,v_y,v_z$ are three real parameters. Note that the explicit form of real Dirac Hamiltonian can be different by a unitary transformation. Here we  consider  $\sigma_0 e^{-\mathrm{i}\frac{\theta \tau_z }{2}}$ to be such unitary transformation, and thus the targeted $k\cdot p$ Hamiltonian can be released to the following form:
\begin{equation}\label{rd-H-1}
    \begin{aligned}
    &H_{RD}(\mathbf{q})=v_z q_z \sigma_z\otimes\tau_0+ v_x q_x \sigma_x\otimes(\cos \theta \tau_x-\sin \theta \tau_y) \\
    &+v_y q_y \sigma_x\otimes\tau_z.
    \end{aligned}
\end{equation}
Then requiring  the linear four band $k\cdot p$ models to take exactly the same form as Eq. \ref{rd-H-1},  we find several MSGs whose  high-symmetry points could  realize the real Dirac fermions,   as listed in Table \ref{tab:rd}.

Interestingly, all of high-symmetry points in Table \ref{tab:rd}   own only one co-irrep, thus the real Dirac fermions essentially exist around these points. Then in order to find ideal real Dirac semimetal, one only need to perform first-principles calculations for the materials crystalizing in the MSGs listed in Table  \ref{tab:rd} and check whether the real Dirac points which necessarily exist are close to the Fermi level.

\begin{table}[!bhtp]
\begin{tabular}{c|c|c|c}
  \hline\hline
  Type& MSG & $\mathbf{k}^*$ name & $\mathbf{k}^*$ coordinate\\\hline
  IV& 129.421& $M$& $(\frac{1}{2},\frac{1}{2},0)$\\\hline
  IV& 129.421& $A$& $(\frac{1}{2},\frac{1}{2},\frac{1}{2})$\\\hline
  IV& 129.422& $M$& $(\frac{1}{2},\frac{1}{2},0)$\\\hline
  IV& 130.433& $M$& $(\frac{1}{2},\frac{1}{2},0)$\\\hline
  IV& 130.434& $M$& $(\frac{1}{2},\frac{1}{2},0)$\\\hline
  IV& 136.505& $A$& $(\frac{1}{2},\frac{1}{2},\frac{1}{2})$\\\hline
  IV& 137.517& $M$& $(\frac{1}{2},\frac{1}{2},0)$\\\hline
  IV& 137.518& $M$& $(\frac{1}{2},\frac{1}{2},0)$\\\hline
  IV& 138.529& $M$& $(\frac{1}{2},\frac{1}{2},0)$\\\hline
  IV& 138.530& $M$& $(\frac{1}{2},\frac{1}{2},0)$\\
  \hline\hline
\end{tabular}\caption{The list of concrete $k$ points which essentially host real Dirac fermions in corresponding MSGs. The MSG names are given in the BNS notation. The SOC is considered. The four columns correspond to type of MSG, name of $k$ point and coordinate of $k$ point, respectively.}\label{tab:rd}
\end{table}

\noindent \textbf{Discussion and Perspective}\\
As a general method, symmetry analysis plays an important role in many areas of natural science, thus  our explicit tabulations  in SM: Part I \cite{SM} of all representation matrices  will be very useful in many aspects.  Using our exhaustive $k\cdot p$ elementary blocks in SM: Part III \cite{SM}, one could quickly construct any $k\cdot p$ model.   The obtained $k\cdot p$ model could then be applied to investigate the electronic structure, transport behavior, response to optical, magnetic and strain field, etc. efficiently.  One can also discretize the  obtained $k\cdot p$ model to   a simple lattice model which shares the same low-energy physics for later complex calculations.  It is worth pointing out that our results  are beyond the scope of conventional applications of $k\cdot p$ models, for they can be used to deal with any number of bands and high-order expansions.  $k\cdot p$ models around  $k$ points in a mainfold can also be constructed and compared.

Our strategy of constructing all $k\cdot p$ models can be applied to impose more symmetry constraints which are beyond  those in MSGs, such as particle-hole symmetry, sublattice symmetry, dual symmetry etc. It can also be applied to the exhaustive constructions of effective models in real space, such as  tight-binding models and spin models. Moreover, realizing those theoretical models with proposed novel properties in crystalline materials has attracted broad interest. The successful examples include Dirac \cite{Na3Bi, Cd3As2, SMYoung}, Weyl \cite{Weyl, TaAs-1, TaAs-2} and Majorana fermions \cite{Qi-RMP}.  Besides, there are still many exotic proposals, such as Rarita-Schwinger particle \cite{RS-1, RS-2} with possible superluminality \cite{RS-3} and  unconventional topological phases \cite{rmp-chiu, unlocking} suggested by various theories, have no materials realizations. One can directly compare the low energy models for them with those constructed from our exhaustive list of elementary  $k\cdot p$ blocks to find candidate MSGs and then based on MSGs, searching for materials is very convenient.     We expect our work could bring  tighter connection between condensed matter physics with other fields like high energy physics in future.

\noindent \textbf{Acknowledgments} We were supported by the National Key R\&D Program of China (Grants No. 2017YFA0303203 and No. 2018YFA0305704), the National Natural Science Foundation of China (NSFC Grants No. 11525417, No. 11834006, No. 51721001, and No. 11790311) and the excellent program at Nanjing University. X.W. also acknowledges the support from
the Tencent Foundation through the XPLORER PRIZE. F.T. was supported by the Fundamental Research Fund for the
Central Universities (No. 14380144) and thanks Prof. Dingyu Xing and Prof. Baigen Wang for their kind and substantial support on scientific research.\\

\textit{Note added.-}During the final editing of our paper, we realized a preprint \cite{yao}  which obtained a complete list of all possible particles in time reversal-invariant systems through symmetry analysis on $k\cdot p$ Hamiltonians around all possible band touchings in special $k$ points.

\bibliography{kp}

\begin{widetext}

\ \\
\newpage

\newpage
\begin{appendix}

{\centering\textbf{Appendix}}

Organization of the Appendix:\\

\noi{\ref{app:msg}} We briefly describe four types of magnetic space groups (MSGs), the classification of special $k$ points in the Brillouin zone (BZ) by point group (unitary), and the computation of the irreducible representations (irreps), $\alpha$,  of the unitary part of the little group: not only the character table $\chi^\alpha$ but also the representation matrices $D^\alpha$ (necessary for the symmetry analysis of $k\cdot p$ Hamiltonians). \\

\noi{\ref{app:Aonirreps}} We then discuss the effect of antiunitary symmetry on the irreps of the unitary part of the little group, which leads to  the co-irreps in the form of $\{\alpha\}$, $\{\beta,\beta\}$  or $\{\gamma,\delta\}$. The calculation of the matrix representation of the antiunitary symmetry operation is also described.\\

\noi{\ref{app:kpAppendix}} We describe the method of constructing any symmetry-allowed $k\cdot p$ Hamiltonian in detail.\\

 \noi{\ref{app:detail}} We show the details of constructing the first-order eight-band Kane model and examples of constructing Dirac models in graphene and the  $k\cdot p$ model that could realize the Hopf-link nodal structure.\\

\section{Magnetic space groups, the classification of special $k$ vectors and  the calculations of irreps $\alpha$}\label{app:msg}

The 1651 MSGs are intimately related with 230 space groups (SGs) \cite{bradley}. In the Belov-Neronova-Smirnova (BNS) notation \cite{bns}, each MSG is assigned a name denoted by $X.Y$ where $X$ denotes the parent SG  from which the MSG can be generated through multiplying  time-reversal operator $\Theta$ to: all SG operations in type II MSGs; half SG operations in type III MSGs; all SG operations in type IV MSGs but with an additional fractional translation.  $Y$ is the sequence number of the MSGs in the same crystal family: triclinic, monoclinic, orthorhombic, tetragonal, trigonal and hexagonal, cubic families \cite{bradley}.   Thus any MSG, denoted by $G$, could be written  in the following form of a summation of co-sets:
\begin{equation}\label{app:msg1}
    G=\left\{\begin{array}{cc}
    \sum_{i=1}^{n_0}\{p_i|\mathbf{t}_i\}T, & \text{for} \quad \text{230 type I MSGs}\\
    \sum_{i=1}^{n_0}\{p_i|\mathbf{t}_i\}T+\Theta\sum_i^{n_0}\{p_i|\mathbf{t}_i\}T, & \text{for} \quad \text{230 type II MSGs}\\
    \sum_{i=1}^{n_0}\{p_i|\mathbf{t}_i\}T+\Theta R_{III}\sum_i^{n_0}\{p_i|\mathbf{t}_i\}T, & \text{for} \quad \text{674 type III MSGs}\\
    \sum_{i=1}^{n_0}\{p_i|\mathbf{t}_i\}T+\Theta R_{IV}\sum_i^{n_0}\{p_i|\mathbf{t}_i\}T, & \text{for} \quad \text{517 type IV MSGs}
    \end{array}\right.
\end{equation}
where  $\{p_i|\mathbf{t}_i\}$ is SG operation in Seitz notation, $n_0$ is the order of the unitary point group $G_0$, and $T$ is the translation group which is the same as SG $X$.  In Eq. \ref{app:msg1},  the point part in the SG  operation $R_{III}$ cannot be identity while the SG operation $R_{IV}$ is just a factional translation.  \textit{We follow the convention which is adopted in Ref. \cite{bradley} throughout this work}. In the Supplementary Material  \cite{SM}: Part I, we show the translation group $T$ by the name of Bravais lattice, for example, $\Gamma_o^f$ represents  the face-centered orthorhombic lattice. As explicitly  given in  Ref. \cite{bradley}, the Cartesian  coordinates of the primitive lattice basis vectors of $\Gamma_o^f$ are:
\begin{equation}\label{app:a1a2a3}
    \left\{
    \begin{array}{c}
    \mathbf{a}_1=(0,b/2,c/2)\\
     \mathbf{a}_2=(a/2,0,c/2)\\
      \mathbf{a}_3=(a/2,b/2,0)\\
    \end{array}
    \right.
\end{equation}
where $a,b,c$ are lattice parameters. Note that a specific Cartesian basis $O$-$xyz$ is chosen not only to express the primitive lattice basis vectors, but also defining the point operations, for example,  $C_{4z}^+$ represents the rotation by  $\pi/2$ around $z$-axis. In SM: Part I \cite{SM}, we show all the MSG operations in the form of $\{p,\{x,y,z\}\}$ where $p$ is the point part such as $C_{4z}^+$ and $\{x,y,z\}$ denotes the translation whose basis vectors are the primitive lattice basis vectors. As in the type II/III/IV MSGs, some operations containing $\Theta$, thus being antiunitary, are printed in red color in SM:Part I \cite{SM}.

For each kind of Bravais lattice, there is a corresponding reciprocal lattice from which there is a corresponding Brillouin zone (BZ).  Sometimes several kinds of BZs should be dinstinguished: For some Bravais lattices, the shape of BZ depends on the lattice parameters \cite{bradley}. For example, with respect to the   face-centered orthorhombic lattice $\Gamma_o^f$, there are four kinds of BZs \cite{bradley} corresponding to four settings: $(a)1/a^2<1/b^2+1/c^2$ and $1/b^2<1/c^2+1/a^2$ and $1/c^2<1/a^2+1/b^2$; $(b)1/c^2>1/a^2+1/b^2$;$(c)1/b^2>1/a^2+1/c^2$;$(d)1/a^2>1/b^2+1/c^2$. Our notation of settings: 1,2,3,4 correspond to $a,b,c,d$ in Ref. \cite{bradley}, respectively.

Though there are infinite number of $k$ points in BZ for infinite crystals, they can be classified by symmetries.  And for $k$ points with the same symmetry, it is enough to analyze only one representative of them. For example, all the $k$ points in a high-symmetry line  own the same symmetry such as $C_n$ and $C_{nv}$ point groups in Schoenflies' notation and the corresponding little groups are also the same, thus the irreps or co-irreps are the same so that the symmetry-constraints on the $k\cdot p$ Hamiltonians should also be the same.  Besides, for symmetry-related $k$ points, their $k\cdot p$ Hamiltonians are also symmetry-related, so that we only need to consider one of the $k\cdot p$ Hamiltonians around symmetry-related $k$ vectors.  In order to exhaustively consider all special $k$ points with no omissions,  we use the follow strategy for classifying $k$ points in the BZ:\\

1. Use the point group $G_b$ of the Bravais lattice to classify the $k$ points in the BZ into special points including  high-symmetry points, high-symmetry lines and high-symmetry planes  based on their point groups. These have been listed in Ref. \cite{bradley} except the high-symmetry planes.   We  obtain the  high-symmetry planes from the surfaces of the irreducible BZs, with the names being $P1$,$P2$,$P3,\ldots$ while their explicit coordinates are given as shown in SM: Part I \cite{SM}. The coordinates of high-symmetry points are three definite numbers, such as $0,1/2,1/4,1/3,\ldots$ and the coordinates for high-symmetry lines and planes  contain 1 and 2 variables,  $hsl\alpha$ and $\{u,v\}$ respectively;\\

2. For each of these $k$ points, denoted as $k_i$, use $G_b$ to obtain all other related $k$ points, thus obtaining a $k$ star: $K_i=\{k_i,k_i',k_i'',\ldots\}$;\\

3. Then consider the real point group $G_0$ (unitary) of a given MSG which must be a subgroup of $G_b$.  $G_0$ could split the parent $k$ star $K_i$ into several real $k$ stars since it is possible that some pair of $k$ vectors in $K_i$ cannot be related by any operation in $G_0$: $K_i$ is split as $K_i=\{K_i^1,K_i^2,\ldots\}$. Each vector in the real $k$ star $K_i^j$ could be related by an element in $G_0$ with another vector in the same real $k$ star, while $G_0$ could not relate different real $k$ stars.\\

Based on such strategy we obtain all real stars for all MSGs. We show all the coordinates of representative $k$ vectors in the real stars in  SM: Part I \cite{SM}. It is still possible that, for example,  some high-symmetry point   would own the same little group as the neighboring high-symmetry line, thus only analyzing  the high-symmetry line is enough, but we still consider them separately. Besides, some real star $K_i^j$ in a parent star $K_i$ could be related by another real star $K_i^{j'}$ in the same parent star through an antiunitary symmetry, while for this case, we still consider them separately.

Given any point $\mathbf{k}$ in an MSG $G$, the little group $G(\mathbf{k})$ contains elements in $G$ which leave $\mathbf{k}$ invariant up to a reciprocal lattice vector. Though $G(\mathbf{k})$ itself is an MSG, the irreps or co-irreps could be found as follows: firstly using the central extension to transform the unitary part $H(\mathbf{k})$ of $G(\mathbf{k})$ to a finite group, whose irreps can be obtained much easier, then if there exists antiunitary symmetry, we  consider the effect of the possible antinuitary symmetry $A$ on the irreps of $H(\mathbf{k})$ in addition. As a matter of fact, similar idea also applies to the construction of elementary $k\cdot p$ blocks.  Firstly, let's introduce how to construct the central extension \cite{bradley}. Define the projective representation  $\Delta$ of $H(\mathbf{k})$ as:
\begin{equation}\label{app:pro}
    \Delta(O_i)=e^{\mathrm{i}\mathbf{k}\cdot \mathbf{t}_{O_i}}D(O_i), O_i\in H(\mathbf{k}),
\end{equation}
where $\mathbf{t}_{O_i}$ denotes the translation part of $O_i$ and $\mathrm{i}^2=-1$. It is easy to find that $\Delta(O_i)$ only depends on the point part of $O_i$, denoted by $p_{O_i}$. Thus $\Delta$ transforms the infinite group $H(\mathbf{k})$ to a (projective) group with  finite elements of order $g_k$, which is the number of elements in the point group of $H(\mathbf{k})$. However, $\Delta$ do not constitute a representation since,
\begin{equation}\label{app:pro-mul}
    \Delta(O_i)\Delta(O_j)=e^{-\mathrm{i}\mathbf{G}_{ij}\cdot \mathbf{t}_{O_j}}\Delta(O_iO_j),
\end{equation}
where $\mathbf{G}_{ij}=p_{O_i}^{-1}\mathbf{k}-\mathbf{k}$ is a reciprocal lattice vector.  To overcome such problem, the central extension \cite{bradley} can be used to construct a new finite group, denoted by $H(\mathbf{k})^*$ whose order is $g_k \lambda$ where $\lambda$ is an integer. $\lambda$ is actually at most 4 for 1651 MSGs. The element in $H^*(\mathbf{k})$ can be written as $(p_{O_i},\eta)$ where $\eta=0,1,\ldots,\lambda-1$. By calculating all  phases $e^{-\mathrm{i}\mathbf{G}_{ij}\cdot \mathbf{t}_{O_j}}$ and  setting them to be equal to $e^{\mathrm{i}2\pi a_{ij}/\lambda}$ ($a_{ij}$ is an integer) we can obtain the least value of $\lambda$. The multiplication of two elements in $H^*(\mathbf{k})$ can be found to be as follows:
\begin{equation}\label{app:pro-mul-cen}
    (p_{O_i},\eta)(p_{O_j},\eta')=(p_{O_i}p_{O_j},\mod(\eta+\eta'+a_{ij},\lambda)),
\end{equation}
thus the central extension $H^*(\mathbf{k})$ is  closed.

Then after constructing the multiplication table from Eq. \ref{app:pro-mul-cen}, the irreps  for $H^*(\mathbf{k})$ (denoted by $a$): character table $\chi^a$ as well as irrep matrices ${D^*}^a$ can be found. Only some of these irreps are those for $H(\mathbf{k})$, satisfying: $(D^*)^a((O_i,\eta))=\Delta^\alpha(O_i)e^{\mathrm{i}2\pi \eta/\lambda}$. We present all the projective matrices for $H(\mathbf{k})$: $\Delta^\alpha(h), h\in H(\mathbf{k})$ in  SM: Part I \cite{SM}. Here we use $\alpha$ to denote the irreps of $H(\mathbf{k})$. $\alpha$  takes $1,2,3\ldots$ whose dimension is $d_\alpha$. Note that the ordering of $\alpha^,s$ is arbitrary, and we do not care too much on the names of these irreps, since we directly and explicitly list their irrep matrices.

\section{Effect of antiunitary symmetry}\label{app:Aonirreps}
Then we consider the effect of the antiunitary symmetry $A$ on the irreps of $H(\mathbf{k})$. Denote $A$ by $\{\beta_a|\boldsymbol{\tau}_a\}\Theta$ and the point part is $A_0=\beta_a$. Let's denote the basis vectors of irrep $\alpha$ by $\{\Psi^\alpha_s\}$ where $s=1,2,\ldots,d_\alpha$ which are also the energy eigenstates simultaneously. The irrep matrices for such basis vectors are just those as listed in SM: Part I \cite{SM}.  $\{A\Psi^\alpha_s\}$ thus would also be an irrep of $H(\mathbf{k})$, whose representation matrices can be found by the following fomula:
\begin{equation}\label{app:Apsi}
    O_i A\Psi^\alpha_s=A O'_i \Psi^\alpha_s={D^\alpha(O'_i)_{s's}^*}A\Psi^\alpha_{s'},
\end{equation}
where Einstein summation rule has been adopted on $s'$ and $O'$ can be found by $O'=A^{-1}O_i A \in H(\mathbf{k})$.  From the characters of  ${D^\alpha(O'_i)_{s's}^*}$, it is easy to know which irrep  the basis vectors: $\{A\Psi^\alpha_s\}$ belongs to.

If the irrep of   $\{A\Psi^\alpha_s\}$ is found to be still $\alpha$,  $\{A\Psi^\alpha_s\}$ may span the same Hilbert space spanned by $\{\Psi^\alpha_s\}$ or orthogonal to the space spanned by $\{\Psi^\alpha_s\}$.
Assume that  $\{A\Psi^\alpha_s\}$ span the same Hilbert space of  $\{\Psi^\alpha_s\}$, thus,  they are up to a unitary transformation matrix, denoted by $U'$ so that ${D^\alpha(O'_i)^*}={U'}^\dag D^{\alpha}(O_i) U'$ where $\{A\Psi^\alpha_s\}=\{\Psi^\alpha_s\}U'$. Here $\{A\Psi^\alpha_s\}=\{\Psi^\alpha_s\}U'$ means,
\[ A\Psi^\alpha_s=\Psi^\alpha_{s's}U'_{s's},\]
from which we know that:
 \[A^2\Psi^\alpha_s=\Psi^\alpha_{s's}(U'{U'}^*)_{s's}.\]

  Since $D^\alpha(A^2)$ is already known ($\pm A^2\in H(\mathbf{k})$ where $\pm$ is originated from that $\Theta^2=1$ or $-1$ for single-valued and double-valued representations, respectively). We should check if $U'$ satisfies $U'U'^*=D^\alpha(A^2)$, and if it is true, it corresponds to case 2 as shown in Eq. \ref{app:herring}.  Thus the assumption that  $\{A\Psi^\alpha_s\}$ span the same Hilbert space of  $\{\Psi^\alpha_s\}$ is right. The representation of $A$ is  $u_\alpha \mathcal{K}$ which is equal to $U' \mathcal{K}$ (the co-irrep is denoted by $\{\alpha\}$), otherwise, the space spanned by $\{A\Psi^\alpha_s\}$ is orthogonal to  that spanned by $\{\Psi^\alpha_s\}$ (case 1) and the representation matrix of $A$ is obtained by:
 \begin{equation}\label{app:repA-2}
    \left(\begin{array}{cc}
    0& u'_{\alpha}\\
    u_{\alpha}& 0
    \end{array}\right)\mathcal{K},
  \end{equation}
in the basis of $\{\Psi^\alpha_s\}\oplus A\{\Psi^\alpha_s\}U'^\dag$   where $u_\alpha=U'$ and $u'_{\alpha}=D^\alpha(A^2){U'}^\top$. In the basis of $\{\Psi^\alpha_s\}\oplus A\{\Psi^\alpha_s\}U'^\dag$, any element $h$ in $H(\mathbf{k})$ is $D^\alpha(h)\oplus D^\alpha(h)$.  And the corresponding co-irrep is represented by $\{\alpha,\alpha\}$ which actually labels the energy level with degeneracy being $2d_{\alpha}$.

If $\{A\Psi^\alpha_s\}$ is found to belong to a different irrep (case 3), $\alpha'\ne\alpha$, the space spanned by $\{A\Psi^\alpha_s\}$ must be orthogonal to that spanned by $\{\Psi^\alpha_s\}$, doubling the degeneracy, and each energy level should be labeled by the co-irrep denoted by $\{\alpha,\alpha'\}$.  For this case,  it is possible that ${D^\alpha(O'_i)^*}$ may not be equal to $D^{\alpha'}(O_i)$ but there must exist a unitary matrix  $U''$ so that ${D^\alpha(O'_i)^*}={U''}^\dag D^{\alpha'}(O_i) U''$. The unitary matrix can be found by the standard method of projection operator in group theory \cite{bradley}. In the basis of $\{\Psi^\alpha_s\}\oplus \{A\Psi^\alpha_s\}{U''}^\dag$, the representation of $h\in H(\mathbf{k})$ would be simply $D^\alpha(h)\oplus D^{\alpha'}(h)$ while the representation of $A$ would become:
  \begin{equation}\label{app:repA-1}
    \left(\begin{array}{cc}
    0& u_{\alpha'}\\
    u_{\alpha}& 0
    \end{array}\right)\mathcal{K},
  \end{equation}
  where $u_\alpha={U}''$ and $u_{\alpha'}=D^\alpha(A^2){U''}^\top$.

Thus we obtain the matrix representations of $A$ for all co-irreps of $G(\mathbf{k})$, and we show them explicitly in SM: Part I \cite{SM}. In fact, cases 1,2,3 can be found directly according to the Herring rule as shown below \cite{bradley}:
\begin{equation}\label{app:herring}
    \left\{\begin{array}{cc}
    \text{case 1:}& \frac{1}{g_k}\sum_{h\in H(\mathbf{k})/T} \chi^\alpha((Ah)^2)=-1\\
        \text{case 2:}& \frac{1}{g_k}\sum_{h\in H(\mathbf{k})/T} \chi^\alpha((Ah)^2)=1\\
            \text{case 3:}& \frac{1}{g_k}\sum_{h\in H(\mathbf{k})/T} \chi^\alpha((Ah)^2)=0,\\
    \end{array}\right.
\end{equation}
in which $g_k$ is the number of cosets $H(\mathbf{k})/T$ while $(Ah)^2=\Theta^2(\{\beta_a|\boldsymbol{\tau}_a\}h)^2=(\pm)(\{\beta_a|\boldsymbol{\tau}_a\}h)^2$  for double-valued (-) or single-valued (+) representations, respectively. In summary, $A$ would cause three possibilities from irreps of $H(\mathbf{k})$ as $\{\alpha\}, \{\beta,\beta\}$ and $\{\gamma,\delta\}$ where $\gamma\ne\delta$. The representation matrices for these three cases are as follows:
\[u_\alpha\mathcal{K}, \left(\begin{array}{cc}0& u'_{\beta}\\u_\beta& 0\end{array}\right)\mathcal{K}, \left(\begin{array}{cc}0& u_{\delta}\\u_{\gamma}&0\end{array}\right)\mathcal{K},\]
respectively.

\section{Method of constructing any symmetry-allowed $k\cdot p$ Hamiltonian}\label{app:kpAppendix}
In this section, we describe how to construct a $k\cdot p$ Hamiltonian, emphasizing a general symmetry method  useful to obtain a $k\cdot p$ data set which can be directly  consulted for a concrete case.  Let's consider a special $k$ point in the BZ, denoted by $\mathbf{k}^*$. $\mathbf{k}^*$ can be high-symmetry point or lie in a high-symmetry line/plane. When it lies in a high-symmetry line/plane, the parameters in the $k\cdot p$ Hamiltonian thus depend on $\mathbf{k}^*$ as it is changed in the high-symmetry line/plane.  The $k\cdot p$ Hamiltonian can be constructed when knowing the participating energy levels (and thus the corresponding Bloch eigenstates) at $\mathbf{k}^*$:
 \begin{equation}\label{energylevels}
    E_{1}\le E_{2}\le E_{3}\le\ldots E_{n}\le\ldots\le E_{N_b},n=1,2,\ldots,N_b,
 \end{equation}
where $N_b$ is the total number of participating energy levels. The corresponding Bloch eigenstates are  denoted by:
 \begin{equation}\label{eigenstates}
 \psi_{1,1},\psi_{1,2},\ldots,\psi_{1,d_1};\psi_{2,1},\ldots,\psi_{2,d_2};\ldots;\psi_{N_b,1},\ldots,\psi_{N_b,d_{N_b}},
 \end{equation}
where $d_n$ denotes the degeneracy of the energy  $E_n$ while  $\psi_{n,s}$ represents the $s$-th Bloch eigenstate for the $n$-th energy level considered.  The central idea of constructing a symmetry allowed $k\cdot p$ Hamiltonian matrix is: the corresponding Bloch eigenstates of the considered energy levels act as the basis kets to represent the lattice Hamiltonian as a matrix, $\mathcal{H}(\mathbf{q})(\mathbf{q}=\mathbf{k}-\mathbf{k}^*)$ which is subject to some constraints given by $D(g)$, namely the representation matrices of the little $G(\mathbf{k}^*)$ in the basis Eq. \ref{eigenstates}:
\begin{equation}\label{symm-constraints}
\begin{aligned}
& D(g) \mathcal{H}(g_0^{-1} \mathbf{q}) D(g)^\dag=\mathcal{H}(\mathbf{q}), g\in H(\mathbf{k}^*),\\
& D(g) \mathcal{H}(-g_0^{-1} \mathbf{q}) D(g)^\dag=\mathcal{H}(\mathbf{q}), g\in G(\mathbf{k}^*)\setminus H(\mathbf{k}^*),
\end{aligned}
\end{equation}
where $g_0$ is the point part of $g$ and $H(\mathbf{k}^*)$ is the unitary part of $G(\mathbf{k}^*)$. It is worth mentioning that we need not arrange the basis kets according to the ascending ordering of energies. Note that when $g\in G(\mathbf{k}^*)\setminus H(\mathbf{k}^*)$, namely, $g\in G(\mathbf{k}^*)$ but $g\notin H(\mathbf{k}^*)$, $g$ can be written by $\{\beta_{a'}|\boldsymbol{\tau}_{a'}\}\Theta$ and the operation on $\mathbf{q}$ contains two parts: one is from the point part $g_0=\beta_{a'}$ namely, $(g_0^{-1}\mathbf{q})$ while the other is time-reversal $\Theta$ (thus $-\mathbf{q}$), so finally, we obtain $-g_0^{-1}\mathbf{q}$.    Eq. \ref{symm-constraints}  can be solved given $D(g)$  in concrete situations.  Note that each energy level at $\mathbf{k}^*$ could be attributed by some (co-)irrep of $G(\mathbf{k}^*)$ (with no accidental degeneracy). We can thus require that the transformation matrices for $\{\psi_{n,s}\}_{s=1}^{d_n}$ in $G(\mathbf{k}^*)$ are exactly the same as those matrices for the (co-)irrep $\xi_n$, ( we use $\xi_n$ to denote the (co-)irrep of the $n$-th energy level). Hence
\begin{equation}\label{rep}
\begin{aligned}
 D(g)=\bigoplus_{n}D^{\xi_n}(g), g\in G(\mathbf{k}^*),\\
\end{aligned}
\end{equation}
where $D^{\xi_n}$ represents the (co-)irrep matrix of $\xi_n$, and note that when $g$ is antiunitary, $D^{\xi_n}(g)$ contains an additional  complex conjugate operation: $\mathcal{K}$ . Note that the co-irrep matrices are all listed explicitly in SM: Part I \cite{SM}: only be careful that the unitary operations are represented by projective representation matrices as in Eq. \ref{app:pro}.

 Hence, Eq. \ref{symm-constraints} can be written by after organizing the $k\cdot p$ Hamiltonian $\mathcal{H}(\mathbf{q})$ into blocks $\mathcal{H}_{nn'}$:
\begin{equation}\label{symm-constraints-1}
\begin{aligned}
& D(g)^{\xi_n} \mathcal{H}(g_0^{-1} \mathbf{q})_{n,n'} {D(g)^{\xi_{n'}}}^\dag=\mathcal{H}(\mathbf{q})_{nn'}, g\in H(\mathbf{k}^*),
\end{aligned}
\end{equation}
and
\begin{equation}\label{symm-constraints-2}
\begin{aligned}
& D(g)^{\xi_n}\mathcal{H}(-g_0^{-1} \mathbf{q})_{n,n'} {D(g)^{\xi_{n'}}}^\dag=\mathcal{H}(\mathbf{q})_{nn'}, g\in G(\mathbf{k}^*)\setminus H(\mathbf{k}^*),
\end{aligned}
\end{equation}
where $\mathcal{H}_{n,n'}$ denotes the block matrix labeled by the row index $n$ and column index $n'$.

Next we firstly describe how to find all solutions satisfying both Eqs. \ref{symm-constraints-1}. We first expand $\mathcal{H}_{nn'}(\mathbf{q})$ as follows:
\begin{equation}\label{kpexpand}
    \mathcal{H}_{nn'}(\mathbf{q})=\sum_{L}\mathcal{H}_{L,nn'}(\mathbf{q}),L=0,1,2,\ldots,
\end{equation}
where $\mathcal{H}_{L,nn'}(\mathbf{q})$ contains the $L$-order polynomials of $\mathbf{q}$. It is straightforward that any $L$-order polynomials can be expanded  on the following basis:
\begin{equation}\label{qqq}
    \{q_x^{n_x}q_y^{n_y}q_z^{n_z},n_x+n_y+n_z=L\},
\end{equation}
which have  $\frac{(L+1)(L+2)}{2}$ polynomials.  Such basis obviously constitutes the basis of the point group $H_0=\{h_0|h\in H(\mathbf{k}^*)\}$ where $h_0$ denotes the point part of $h$, so that we can analyze Eq. \ref{symm-constraints-1} for each $L$ separately. However, we could further make simplifications. We observe that $L$-order polynomials must constitute a basis for  a representation of $SO(3)$ group. In next subsection, we would describe a new basis polynomials derived from the basis of irrep of $SO(3)$, denoted by the angular momentum quantum number  $l=0,1,2,3,4,\ldots$

\subsection{Basis functions for polynomials of $\mathbf{q}$}\label{q-basis}
Here we would directly  give the expressions for the polynomial basis functions which are derived from spherical harmonics. For expansion order $L=0,1,2\ldots$,  the most natural polynomial basis functions might be $\{q_x^{n_x}q_y^{n_y}q_z^{n_z}, n_x+n_y+n_z=L\}$ of which the total number of basis functions is $\frac{(L+1)(L+2)}{2}$ as mentioned above. However, in this work, we take a different choice related with spherical harmonics. From $Y_{lm}(\theta,\phi)$ ($\theta=\arccos\frac{q_z}{q},\phi=\arctan\frac{q_y}{q_x}$), through unitary transformation and multiplying the results by $q^l$, one could obtain the real basis functions as shown below in Table \ref{tab2}.   In the following $l=0,1,2,3,4$ are represented by $S,P,D,F,G$, respectively. For example, for $l=2$, there should be $5$ basis functions, denoted as $D_1,D_2,D_3,D_4,D_5$. Note that the argument $\mathbf{q}$ has been omitted for simplicity. For expansion order $L$, we should consider a set of basis functions for polynomials  originated from  $l^,s$: $l=L,L-2,L-4,\ldots$ and the corresponding basis functions are those for $l$ (shown in Table \ref{tab2}) multiplied by $q^{L-l}$ (note that $q^{L-l}$ wouldn't change the transformation properties of the basis functions). It is easy to check that  such new choice actually corresponds to decompose  $\{q_x^{n_x}q_y^{n_y}q_z^{n_z}, n_x+n_y+n_z=L\}$ to the basis functions of irreps labeled by $l$ of $SO(3)$ group. The dimension is conserved:
\[\frac{(L+1)(L+2)}{2}=\sum_{l=L,L-2,L-4,\ldots,l\ge0}(2l+1).\]
\begin{table}[!thbp]
\begin{tabular}{c|c|c|c|c}
$l=0$& $l=1$& $l=2$& $l=3$& $l=4$\\\hline
$S_1=1$&  $P_1=q_x$ &$D_1=\frac{1}{2} \left(q_x^2-q_y^2\right)$&$F_1=\frac{q_x \left(q_x^2-3 q_y^2\right)}{2 \sqrt{6}}$ & $G_1=\frac{1}{4} \left(-6 q_x^2 q_y^2+q_x^4+q_y^4\right)$\\
-& $P_2=q_y$ & $D_2=q_x q_y$& $F_2=-\frac{q_y \left(q_y^2-3 q_x^2\right)}{2 \sqrt{6}}$& $G_2=q_x q_y \left(q_x^2-q_y^2\right)$\\
-& $P_3=q_z$& $D_3=q_x q_z$& $F_3=\frac{1}{2} q_z \left(q_x^2-q_y^2\right)$& $G_3=\frac{q_x q_z \left(q_x^2-3 q_y^2\right)}{\sqrt{2}}$\\
-& -& $D_4=q_y q_z$& $F_4=q_x q_y q_z$& $G_4=-\frac{q_y q_z \left(q_y^2-3 q_x^2\right)}{\sqrt{2}}$\\
-& -& $D_5=-\frac{q_x^2+q_y^2-2 q_z^2}{2 \sqrt{3}}$& $F_5=-\frac{q_x \left(q_x^2+q_y^2-4 q_z^2\right)}{2 \sqrt{10}}$& $G_5=-\frac{\left(q_x^2-q_y^2\right) \left(q_x^2+q_y^2-6 q_z^2\right)}{2 \sqrt{7}}$\\
-& -& -& $F_6=-\frac{q_y \left(q_x^2+q_y^2-4 q_z^2\right)}{2 \sqrt{10}}$& $G_6=-\frac{q_x q_y \left(q_x^2+q_y^2-6 q_z^2\right)}{\sqrt{7}}$\\
-& -& -& $F_7=\frac{2 q_z^3-3 q_z \left(q_x^2+q_y^2\right)}{2 \sqrt{15}}$& $G_7=\frac{q_x q_z \left(4 q_z^2-3 \left(q_x^2+q_y^2\right)\right)}{\sqrt{14}}$\\
-& -& -& -& $G_8=\frac{q_y q_z \left(4 q_z^2-3 \left(q_x^2+q_y^2\right)\right)}{\sqrt{14}}$\\
-& -& -& -& $G_9=\frac{-24 q_z^2 \left(q_x^2+q_y^2\right)+3 \left(q_x^2+q_y^2\right){}^2+8 q_z^4}{4 \sqrt{35}}$
\end{tabular} \caption{Explicit expressions for basis functions of polynomials of $\mathbf{q}$ characterized by $l=0,1,2,3,4$.}\label{tab2}
\end{table}

Hence, the $L$-order basis polynomials in Eq. \ref{qqq}  can be transformed to the following form:
\begin{equation}\label{poly-l}
\{q^{L-l}f^l_{m_f}(\mathbf{q})\}_{l=L,L-2,L-4,\ldots},
\end{equation}
where $f^l_{m_f}$ represent the basis polynomials from $l$ as described above and $m_f$ can take $2l+1$ values.
Let us check that the numbers of basis polynomials for Eqs. \ref{qqq} and \ref{poly-l} are the same  for $L=0,1,2,3,4$ one by one. For $L=0$, Eq. \ref{qqq} and Eq. \ref{poly-l} are both $\{1\}$. For $L=1$, Eqs. \ref{qqq} and \ref{poly-l} are both $\{q_x,q_y,q_z\}$. For $L=2$, Eq. \ref{qqq} is $\{q_x^2,q_y^2,q_z^2,q_xq_y,q_yq_z,q_xq_z\}$ containing 6 polynomials, while for Eq. \ref{poly-l}, $l$ can be $2$ and $0$, thus the total number of polynomials in Eq. \ref{poly-l} is also $6=5+1$,  and these polynomial basis functions are, $\{D_1,D_2,D_3,D_4,D_5,q^2\}$. For $L=3$, Eq. \ref{qqq} is $\{q_x^3,q_y^3,q_z^3,q_x^2q_y,q_y^2q_x,q_x^2q_z,q_z^2q_x,q_y^2q_z,q_z^2q_x,q_xq_yq_z\}$ containing 10 polynomials while in Eq. \ref{poly-l}, $l$ can be $3$ and $1$ so that there are $7+3=10$ polynomial basis functions: $\{F_1,F_2,F_3,F_4,F_5,F_6,F_7,q^2 P_1,q^2 P_2,q^2 P_3\}$. For $L=4$,  Eq. \ref{qqq} contain 12 polynomials as\\ $\{q_x^4,q_y^4,q_z^4,q_x^3q_y,q_y^3q_x,q_x^3q_z,q_z^3q_x,q_y^3q_z,q_z^3q_y,q_x^2q_yq_z,q_y^2q_xq_z,q_z^2q_xq_y,q_x^2q_y^2,q_x^2q_z^2,q_y^2q_z^2\}$, while in Eq. \ref{poly-l}, $l=4,2,0$, so that there are in total $9+5+1=15$ polynomials as
$\{G_1,G_2,G_3,G_4,G_5,G_6,G_7,G_8,G_9,q^2D_1,q^2D_2,q^2D_3,q^2D_4,q^2D_5,q^4\}$.  Since we use Eq. \ref{poly-l} in this work, and $\mathcal{H}_{L,nn'}(\mathbf{q})$ can be  written as the following expansion formally:
\begin{equation}\label{expand-l}
    \mathcal{H}_{L,nn'}(\mathbf{q})=\sum_{l=L,L-2,\ldots;m_f} c_{L,l,m_f} q^{L-l}f^l_{m_f}(\mathbf{q})=\sum_{l=L,L-2,\ldots}q^{L-l}\mathcal{H}_{nn'}^l(\mathbf{q}),
\end{equation}
of which $\mathcal{H}_{nn'}^l(\mathbf{q})=\sum_{m_f}c_{L,l,m_f}f^l_{m_f}(\mathbf{q})$ can be analyzed separately.  After we know $\mathcal{H}_{nn'}^l$, we thus use Eq. \ref{expand-l} to know the $L$-order Hamiltonian $\mathcal{H}_{L,nn'}$.

As shown in Eq. \ref{expand-l}, $\mathcal{H}_{nn'}^{l}(\mathbf{q})$ is expanded on $\{f^l_{m_f}\}$  but the expansion coefficients are still matrices in general. We thus choose a set of matrix basis vectors to expand these matrices which would be described in the following. Before showing the matrix basis vectors, we note that, since $\xi_{n}=\alpha$ if $G(\mathbf{k}^*)=H(\mathbf{k}^*)$ where we use $\alpha,\beta,\gamma,\delta,\ldots$ to denote the irreps of $H(\mathbf{k}^*)$ and if $G(\mathbf{k}^*)>H(\mathbf{k}^*)$ (there is antiunitary symmetry, thus $G(\mathbf{k}^*)=H(\mathbf{k}^*)+A\cdot H(\mathbf{k}^*)$), $\xi_n$ can be in the form of $\{\alpha\}$ or $\{\beta,\gamma\}$ ($\beta$ may be equal to $\gamma$),  to solve Eq. \ref{symm-constraints-1} (we firstly consider Eq. \ref{symm-constraints-1} and then further consider Eq. \ref{symm-constraints-2}),  we only need to find $H^l_{\alpha\beta}$ for all pairs of irreps $(\alpha,\beta)$,  satisfying:
\begin{equation}\label{ab}
    D^\alpha (h) H_{\alpha\beta}^l(h_0^{-1}\mathbf{q}) {D^\beta(h)}^\dag=H_{\alpha\beta}^l(\mathbf{q}), h\in H(\mathbf{k}^*),
\end{equation}
where $H_{\alpha\beta}^l(\mathbf{q})$ is combination of $\{f^l_{m_f}\}$. From $H_{\alpha\beta}^l(\mathbf{q})$, we can quickly obtain $\mathcal{H}^l_{n,n'}(\mathbf{q})$ satisfying Eq. \ref{symm-constraints-1}, by,
\begin{equation}\label{get-l}
    \mathcal{H}^l_{n,n'}(\mathbf{q})=\left\{\begin{array}{cr}
H^l_{\alpha,\beta}(\mathbf{q}),& \text{if}\quad \xi_n=\alpha, \xi_{n'}=\beta \quad\text{or}\quad \xi_n=\{\alpha\}, \xi_{n'}=\{\beta\}\\
\left(\begin{array}{cc}H^l_{\alpha,\beta}(\mathbf{q})&H^l_{\alpha,\gamma}(\mathbf{q})\end{array}\right),& \text{if}\quad \xi_n=\{\alpha\}, \xi_{n'}= \{\beta,\gamma\}\\
\left(\begin{array}{c}H^l_{\alpha,\gamma}(\mathbf{q})\\H^l_{\beta,\gamma}(\mathbf{q})\\
\end{array}\right),& \text{if}\quad \xi_n=\{\alpha,\beta\}, \xi_{n'}= \{\gamma\}\\
\left(\begin{array}{cc}H^l_{\alpha,\gamma}(\mathbf{q})& H^l_{\alpha,\delta}(\mathbf{q})\\H^l_{\beta,\gamma}(\mathbf{q})&H^l_{\beta,\delta}(\mathbf{q})\\
\end{array}\right),& \text{if}\quad \xi_n=\{\alpha,\beta\}, \xi_{n'}= \{\gamma,\delta\}\\
\end{array}\right.
\end{equation}

Our matrix basis vectors are chosen to expand $H^l_{\alpha,\beta}(\mathbf{q})$ as described below.

\subsection{matrix basis}\label{m-basis}

In this work, two kinds of matrix basis sets: $\{\Gamma_\mu^{\alpha\beta}\}$ are chosen as described below:

When $\alpha=\beta$, we chose the Hermitian matrix basis vectors so that ${\Gamma^{\alpha\alpha}_\mu}^\dag=\Gamma^{\alpha\alpha}_\mu$. 
The concrete form of  $\Gamma_\mu^{\alpha\alpha}$ in this case is: The first $d_\alpha$ basis vectors are $\Gamma_\mu^{\alpha\alpha}=E_\mu (\mu\le d_\alpha)$ where $E_\mu$ means that only the $(\mu,\mu)$ entry is 1 and the other entries are zero. Then for $\mu=d_\alpha+1$, $\Gamma^{\alpha\alpha}_\mu=S_{\{1,2\}}$, and $S_{\{1,2\}}$ denotes a symmetry matrix and only $(1,2)$ and $(2,1)$ entries are $\frac{1}{\sqrt 2}$ and the rest entries are zero. Then  for $\mu=d_\alpha+2$, $\Gamma^{\alpha\alpha}_\mu=A_{\{1,2\}}$, and $A_{\{1,2\}}$ is an antisymmetry matrix and only $(1,2)$ and $(2,1)$ entries are $-\frac{\mathrm{i}}{\sqrt 2}$ and $\frac{\mathrm{i}}{\sqrt 2}$, respectively,  and the rest entries are zero. Subsequently, for $\mu=d_\alpha+3$, $\Gamma^{\alpha\alpha}_\mu=S_{\{1,3\}}$ and then $\Gamma^{\alpha\alpha}_{d_\alpha+4}=A_{\{1,3\}}$, and so on. In general, for $i<j$,
\begin{equation}\label{Appendix-eq-5}
\begin{aligned}
&\Gamma^{\alpha\alpha}_{\mu}=S_{\{i,j\}}, \qquad if \qquad \mu=d_\alpha+2(\sum_{a=1}^{i-1} (d_\alpha-a))+2(j-i-1)+1,\\
&\Gamma^{\alpha\alpha}_{\mu}=A_{\{i,j\}}, \qquad if \qquad \mu=d_\alpha+2(\sum_{a=1}^{i-1} (d_\alpha-a))+2(j-i),\\
\end{aligned}
\end{equation}
where $S_{\{i,j\}}$ is a symmetry matrix and only $(i,j)$ and $(j,i)$ entries are $\frac{1}{\sqrt 2}$ and the rest entries are zero, and $A_{\{i,j\}}$ is an antisymmetry matrix and only $(i,j)$ and $(j,i)$ entries are $-\frac{\mathrm{i}}{\sqrt 2}$ and $\frac{\mathrm{i}}{\sqrt 2}$, respectively,  and the rest entries are zero.

When $\alpha\ne\beta$, we choose the following basis set as:
\begin{equation}\label{Appendix-eq-6}
    \Gamma^{\alpha\beta}_\mu=M_{\{i,j\}}, \qquad if \qquad \mu=d_\beta(i-1)+j,
\end{equation}
where $M_{\{i,j\}}$ is the matrix of which only the $(i,j)$ entry is 1 and the rest are zero.  Concretely, $\Gamma^{\alpha\beta}_1=M_{\{1,1\}}$, $\Gamma^{\alpha\beta}_2=M_{\{1,2\}}$, $\Gamma^{\alpha\beta}_3=M_{\{1,3\}}$ and so on.

The above two kinds of matrix basis vectors only depends on the dimension of $d_{\alpha}$ and $d_{\beta}$. We can omit the superscripts $\alpha$ and $\beta$ in these matrix basis vectors since we can always know them in the context.

\subsection{How to obtain elementary $k\cdot p$ blocks: $\{H^l_{\alpha\beta m}\}_m$}
Write $H^l_{\alpha\beta}(\mathbf{q})=\sum_{m}c_m H^l_{\alpha\beta m}(\mathbf{q})$ where $H^l_{\alpha\beta m}(\mathbf{q})$ satisfies Eq. \ref{ab} and can be expressed as,
 \begin{equation}\label{Hab-1}
    H^l_{\alpha\beta m}(\mathbf{q})=\sum_{m_f\mu}c^m_{m_f,\mu}f^l_{m_f}(\mathbf{q})\Gamma^{\alpha\beta}_\mu,
 \end{equation}
and span the solution space of Eq. \ref{ab}.  

To find $H^l_{\alpha\beta m}(\mathbf{q})$,  we first obtain the representation matrices of $H(\mathbf{k}^*)$  in the matrix basis $\{\Gamma^{\alpha\beta}_\mu\}_\mu$ as:
\begin{equation}\label{rep-matrix}
    D^M(h)_{\mu\mu'}=\mathrm{Tr}[{\Gamma^{\alpha\beta}_{\mu}}^\dag D(h)\Gamma^{\alpha\beta}_{\mu'}D(h)^\dag], h\in H(\mathbf{k}^*).
\end{equation}
It is obvious that $D^M(h)$ only depends on the point parts of $h$, namely $h_0$, so that the representation (single-valued) of point group $H_0$ is enough.

We then could find a unitary transformation matrix: $U_{\mu,(\zeta,i_\zeta,s_\zeta)}$. Here $i_\zeta$ denotes the occurrence of irrep $\zeta$ of the point group $H_0$ while $s_\zeta$ denote the basis vectors of $\zeta$. Given $\zeta$ and $i_\zeta$, we have:

\begin{equation}\label{matrix-irep}
D^M_{\mu\mu'}(h_0) U_{\mu',(\zeta,i_\zeta,s_\zeta)}=\sum_{s'_\zeta} D^\zeta(h_0)_{s'_\zeta s_\zeta} U_{\mu,(\zeta,i_\zeta,s'_\zeta)}.
\end{equation}

For the polynomials $\{f^l_{m_f}\}$, we can also find a unitary transformation matrix:  $U_{m_f,(\zeta,j_\zeta,s_{\zeta})}$, satisfying,

\begin{equation}\label{poly-irep}
D^l_{m_fm'_f} U_{m'_f,(\zeta,j_\zeta,s_\zeta)}=\sum_{s'_\zeta} {D^\zeta(h_0)}^*_{s'_\zeta s_\zeta} U_{m_f,(\zeta,j_\zeta,s'_\zeta)},
\end{equation}
where $D^l$ is the representation matrix of $h_0$ in the basis of $\{f^l_{m_f}\}_{m_f}$.
Note that in  Eqs. \ref{matrix-irep} and \ref{poly-irep}  the irrep matrices $D^\zeta(p_0)$ and ${D^\zeta(p_0)}^*$ are conjugate with each other, which is important for  proving Eq. \ref{Hab-2} later.
Thus,
\begin{equation}\label{Hab-2}
H^l_{\alpha\beta m}(\mathbf{q})=\sum_{s_\zeta,\mu,m_f} U_{\mu,(\zeta,i_\zeta,s_\zeta)}\Gamma^{\alpha\beta}_{\mu}U_{m_f,(\zeta,j_{\zeta},s_{\zeta})}f^l_{m_f}(\mathbf{q}),
\end{equation}
where $m=(\zeta,i_\zeta,j_\zeta)$.
This is because,
\begin{equation}\label{Hab-3}
  \begin{aligned}
  &  D^\alpha(h)H^l_{\alpha\beta m}(h_0^{-1}\mathbf{q}){D^\beta(h)}^\dag=\sum_{s_\zeta,\mu,m_f} U_{\mu,(\zeta,i_\zeta,s_\zeta)}[D^\alpha(h)\Gamma^{\alpha\beta}_{\mu}{D^\beta(h)}^\dag]U_{m_f,(\zeta,j_{\zeta},s_{\zeta})}f^l_{m_f}(h_0^{-1} \mathbf{q})=\\
  &\sum_{s_\zeta,\mu',m'_f,\mu,m_f}U_{\mu,(\zeta,i_\zeta,s_\zeta)} (D^M_{\mu'\mu}\Gamma^{\alpha\beta}_{\mu'})U_{m_f,(\zeta,j_{\zeta},s_{\zeta})}[D^l_{m'_f,m_f}f^l_{m'_f}(\mathbf{q})]=\\
  & \sum_{s'_\zeta,s''_\zeta,s_\zeta,\mu',m'_f}\Gamma^{\alpha\beta}_{\mu'}(D^\zeta_{s'_\zeta,s_\zeta}(h_0) U_{\mu',(\zeta,i_\zeta,s'_\zeta)})f^l_{m'_f}(\mathbf{q})({D^\zeta_{s''_\zeta,s_\zeta}(h_0)}^* U_{m'_f,(\zeta,j_\zeta,s''_\zeta)})=\\
  &\sum_{s'_\zeta,s''_\zeta,s_\zeta,\mu',m'_f}(D^\zeta_{s'_\zeta,s_\zeta}(h_0){D^\zeta_{s''_\zeta,s_\zeta}(h_0)}^*)\Gamma^{\alpha\beta}_{\mu'}( U_{\mu',(\zeta,i_\zeta,s'_\zeta)})f^l_{m'_f}(\mathbf{q})( U_{m'_f,(\zeta,j_\zeta,s''_\zeta)})=\\
  &\sum_{s'_\zeta,s''_\zeta,\mu',m'_f}(\delta_{s'_\zeta,s''_\zeta})\Gamma^{\alpha\beta}_{\mu'}( U_{\mu',(\zeta,i_\zeta,s'_\zeta)})f^l_{m'_f}(\mathbf{q})( U_{m'_f,(\zeta,j_\zeta,s''_\zeta)})=\\
    &\sum_{s'_\zeta,\mu',m'_f}\Gamma^{\alpha\beta}_{\mu'}( U_{\mu',(\zeta,i_\zeta,s'_\zeta)})f^l_{m'_f}(\mathbf{q})( U_{m'_f,(\zeta,j_\zeta,s'_\zeta)})=H^{l}_{\alpha\beta m}(\mathbf{q}).
    \end{aligned}
\end{equation}

As in Eq. \ref{Hab-2}, $m=(\zeta,i_\zeta,j_\zeta)$. And we directly give $\{H^l_{\alpha\beta m}\}_m$ in  SM: Part III \cite{SM}, where $m$ can be understood as $1,2,3,\ldots$ For $\alpha=\beta$, we set all of $\{H^l_{\alpha\beta m}\}_m$ to be Hermitian. When $G(\mathbf{k}^*)=H(\mathbf{k}^*)$, namely, these is no antiunitary stmmetry,  $\{H^l_{\alpha\beta m}(\mathbf{q})\}_m$ are enough to determine any $k\cdot p$ Hamiltonian. We list all these matrices in  SM: Part III \cite{SM}: for $\alpha=1,2,\ldots$ and $\beta=\alpha,\alpha+1,\ldots$, for each of $l=0,1,2,3,4$. In the following is an example.

\subsection{An example}\label{app:195}
Consider the single-valued representations of $\Gamma$ point in SG 195 with no time-reversal symmetry (TRS), namely, type-I MSG195.1. As shown in   SM: Part I \cite{SM}, it owns 4  irreps denoted by $1,2,3,4$ whose dimensions are 1,1,1,3 respectively.  Let's construct the $k\cdot p$ Hamiltonian when three energy levels are considered: $E_a<E_b<E_c$ and their irreps are $\xi_a=1,\xi_b=4, \xi_c=4$.  Thus the degeneracies of the energy levels are $d_a=1, d_b=d_c=3$ so the $k\cdot p$ Hamiltonian is  a seven-band model and can be written in the following block matrix:
\begin{equation}\label{example-1-Hkp}
    \mathcal{H}(\mathbf{q})=\left(\begin{array}{ccc}
\mathcal{H}(\mathbf{q})_{aa}& \mathcal{H}(\mathbf{q})_{ab}&  \mathcal{H}(\mathbf{q})_{ac}\\
\mathcal{H}(\mathbf{q})_{ba}& \mathcal{H}(\mathbf{q})_{bb}& \mathcal{H}(\mathbf{q})_{bc} \\
\mathcal{H}(\mathbf{q})_{ca}& \mathcal{H}(\mathbf{q})_{cb}& \mathcal{H}(\mathbf{q})_{cc}
\end{array}
\right),
\end{equation}
where $\mathcal{H}_{aa}$ is $1\times 1$ matrix, $\mathcal{H}_{bb}$, $\mathcal{H}_{bc}$ and $\mathcal{H}_{cc}$ are $3\times 3$ matrices and $\mathcal{H}_{ab}$ and $\mathcal{H}_{ac}$ are $1\times 3$ matrices.
First considering $L=0$.  We thus should consider $l=0$. As shown in  SM: Part III \cite{SM}, $\{H^0_{1,1,m}\}=\{E_1S_1\}$, $\{H^0_{4,4,m}\}=\{\frac{E_1+E_2+E_3}{\sqrt 3}S_1\}$, while $\{H^0_{i,j,m}\}=\{\},\forall i<j$ (here $\{\}$ means that no $k\cdot p$ blocks satisfy the symmetry constraints other than 0), hence,
$\mathcal{H}^0(\mathbf{q})_{aa}=\sum_{m}r^1_m H^0_{1,1,m}=r^1_1 E_1S_1$,
$\mathcal{H}^0(\mathbf{q})_{bb}=\sum_{m}r^2_m H^0_{4,4,m}=r^2_1 \frac{E_1+E_2+E_3}{\sqrt3}S_1$,
$\mathcal{H}^0(\mathbf{q})_{bb}=\sum_{m}r^3_m H^0_{4,4,m}=r^3_1 \frac{E_1+E_2+E_3}{\sqrt3}S_1$,
$\mathcal{H}^0(\mathbf{q})_{ab}=0$,
$\mathcal{H}^0(\mathbf{q})_{ac}=0$,
$\mathcal{H}^0(\mathbf{q})_{bc}=\sum_{m}c^1_m H^0_{4, 4,m}=c^1_1 \frac{E_1+E_2+E_3}{\sqrt3}S_1$, where $E_i$ and $S_1$ can be found in Sec. \ref{m-basis} and \ref{q-basis}, respectively. Hence, the finally form of $\mathcal{H}_0(\mathbf{q})$ is,

\begin{equation}\label{example-1-Hkp-0}
    \mathcal{H}_0(\mathbf{q})=\left(\begin{array}{ccc}
\epsilon_a I_{1}&0&0\\
0& \epsilon_b I_{3}& v I_3\\
0& {v}^* I_3 & \epsilon_c I_{3}
\end{array}
\right),
\end{equation}
where $I_n$ means the $n\times n$ identity matrix and $\epsilon_{a}=r^1_1,\epsilon_{b}=\frac{r^2_{1}}{\sqrt 3},\epsilon_{c}=\frac{r^3_1}{\sqrt3}$ are real while $v=c^1_1$ is complex. This occurs actually before diagonalization to obtain the eigen-energy solutions of $\Gamma$ point, after which the $L=0$ order Hamiltonian is $\mathrm{dia}(E_a,E_b,E_b,E_b,E_c,E_c,E_c)$, or we can set $v=0$ while $\epsilon_{a/b/c}=E_{a/b/c}$ directly.

Then consider $L=1$ and thus $l=1$. As shown in  SM: Part III \cite{SM}, $\{H^1_{1,1,m}\}=\{\}$, $\{H^1_{1,4,m}\}=\{P_1M_{\{1,1\}}+P_2M_{\{1,2\}}+P_3M_{\{1,3\}}\}$ and $\{H^1_{4,4,m}\}=\{P_3S_{\{1,2\}}+P_2S_{\{1,3\}}+P_1S_{\{2,3\}},-P_1A_{\{2,3\}}+P_2A_{\{1,3\}}-P_3A_{\{1,2\}}\}$, thus, we obtain:
$\mathcal{H}^{1}_{aa}(\mathbf{q})=0$, $\mathcal{H}^{1}_{ab}=c^1_1P_1M_{\{1,1\}}+c^1_1P_2M_{\{1,2\}}+c^1_1P_3M_{\{1,3\}}$,
$\mathcal{H}^{1}_{ac}=c^2_1P_1M_{\{1,1\}}+c^2_1P_2M_{\{1,2\}}+c^2_1P_3M_{\{1,3\}}$, and $\mathcal{H}^1_{bb}=r^1_1(P_3S_{\{1,2\}}+P_2S_{\{1,3\}}+P_1S_{\{2,3\}})+r^1_2(-P_1A_{\{2,3\}}+P_2A_{\{1,3\}}-P_3A_{\{1,2\}})$ (here two parameters $r^1_1$ and $r^1_2$ are due to that the number of the corresponding elementary $k\cdot p$ blocks is 2),
$\mathcal{H}^1_{cc}=r^2_1(P_3S_{\{1,2\}}+P_2S_{\{1,3\}}+P_1S_{\{2,3\}})+r^2_2(-P_1A_{\{2,3\}}+P_2A_{\{1,3\}}-P_3A_{\{1,2\}})$,
$\mathcal{H}^1_{bc}=c^3_1(P_3S_{\{1,2\}}+P_2S_{\{1,3\}}+P_1S_{\{2,3\}})+c^3_2(-P_1A_{\{2,3\}}+P_2A_{\{1,3\}}-P_3A_{\{1,2\}})$. Then the first-order $k\cdot p$ Hamiltonian, $\mathcal{H}_1(\mathbf{q})$, is (the parameters have been renamed):

\begin{equation}\label{example-1-Hkp-1}
\begin{aligned}
&    \mathcal{H}_1(\mathbf{q})=\\
&\begin{bmatrix}
\begin{array}{c|c|c}
0I_{2}& v_1 (P_1M_{\{1,1\}}+P_2M_{\{1,2\}}+P_3M_{\{1,3\}})& v_2 (P_1M_{\{1,1\}}+P_2M_{\{1,2\}}+P_3M_{\{1,3\}})\\\hline
h.c.& \begin{aligned}r_1(P_3S_{\{1,2\}}+P_2S_{\{1,3\}}+P_1S_{\{2,3\}})+\\ r_2(-P_1A_{\{2,3\}}+P_2A_{\{1,3\}}-P_3A_{\{1,2\}})\end{aligned} & \begin{aligned}c_1(P_3S_{\{1,2\}}+P_2S_{\{1,3\}}+P_1S_{\{2,3\}})+\\c_2(-P_1A_{\{2,3\}}+P_2A_{\{1,3\}}-P_3A_{\{1,2\}})\end{aligned}\\\hline
h.c.&h.c. &\begin{aligned}r_3(P_3S_{\{1,2\}}+P_2S_{\{1,3\}}+P_1S_{\{2,3\}})+\\r_4(-P_1A_{\{2,3\}}+P_2A_{\{1,3\}}-P_3A_{\{1,2\}})\end{aligned}
\end{array}
\end{bmatrix},
\end{aligned}
\end{equation}
where $\{r_i\}$ are real parameters while $\{c_i\}$,$\{v_i\}$ are all complex parameters.

 Then consider $L=2$, and  $l$ can thus take $2$ and $0$. For $l=0$, from Eq. \ref{example-1-Hkp-0}, the corresponding contribution to $\mathcal{H}_2$ should be:
\begin{equation}\label{example-1-Hkp-3}
   q^2\left(\begin{array}{ccc}
r_5 I_{2}&0&0\\
0& r_6 I_{3}& v_3 I_3\\
0& {v_3}^* I_3 & r_7 I_{3}
\end{array}
\right).
\end{equation}

For $l=2$, as shown in  SM: Part III \cite{SM}, $\{H^2_{1,1,m}\}=\{\}$, $\{H^2_{4,4,m}\}=\{\sqrt{\frac{2}{3}}D_1E_1-\frac{D_1E_2}{\sqrt6}-\frac{D_5E_2}{\sqrt2}-\frac{D_1E_3}{\sqrt6},
-\sqrt{\frac{2}{3}}D_5E_1-\frac{D_1E_2}{\sqrt2}+\frac{D_5E_2}{\sqrt6}+\frac{D_1E_3}{\sqrt2}+\frac{D_5E_3}{\sqrt6},D_2S_{\{1,2\}}+D_3S_{\{1,3\}}+D_3S_{\{2,3\}},
-D_2A_{\{1,2\}}+D_3A_{{1,3}}-D_4A_{\{2,3\}}\}$ and $\{H^2_{1,4,m}\}=\{D_4M_{\{1,1\}}+D_3M_{\{1,2\}}+D_2M_{\{1,3\}}\}$. Hence, we would obtain,

\begin{equation}\label{example-1-Hkp-4}
\begin{aligned}
&    \mathcal{H}^{l=2}(\mathbf{q})=\\
&\begin{bmatrix}
\begin{array}{c|c|c}
0I_{2}& v_4 (D_4M_{\{1,1\}}+D_3M_{\{1,2\}}+D_2M_{\{1,3\}})& v_5 (D_4M_{\{1,1\}}+D_3M_{\{1,2\}}+D_2M_{\{1,3\}})\\\hline
h.c.& \begin{aligned}r_8(\sqrt{\frac{2}{3}}D_1E_1-\frac{D_1E_2}{\sqrt6}-\frac{D_5E_2}{\sqrt2}-\frac{D_1E_3}{\sqrt6})+\\ r_9(-\sqrt{\frac{2}{3}}D_5E_1-\frac{D_1E_2}{\sqrt2}+\frac{D_5E_2}{\sqrt6}+\frac{D_1E_3}{\sqrt2}+\frac{D_5E_3}{\sqrt6})+\\
r_{10}(D_2S_{\{1,2\}}+D_3S_{\{1,3\}}+D_3S_{\{2,3\}})+\\
r_{11}(-D_2A_{\{1,2\}}+D_3A_{{1,3}}-D_4A_{\{2,3\}})
\end{aligned} & \begin{aligned}c_3(\sqrt{\frac{2}{3}}D_1E_1-\frac{D_1E_2}{\sqrt6}-\frac{D_5E_2}{\sqrt2}-\frac{D_1E_3}{\sqrt6})+\\ c_4(-\sqrt{\frac{2}{3}}D_5E_1-\frac{D_1E_2}{\sqrt2}+\frac{D_5E_2}{\sqrt6}+\frac{D_1E_3}{\sqrt2}+\frac{D_5E_3}{\sqrt6})+\\
c_{5}(D_2S_{\{1,2\}}+D_3S_{\{1,3\}}+D_3S_{\{2,3\}})+\\
c_{6}(-D_2A_{\{1,2\}}+D_3A_{{1,3}}-D_4A_{\{2,3\}})
\end{aligned} \\\hline
h.c.&h.c. &\begin{aligned}r_{12}(\sqrt{\frac{2}{3}}D_1E_1-\frac{D_1E_2}{\sqrt6}-\frac{D_5E_2}{\sqrt2}-\frac{D_1E_3}{\sqrt6})+\\ r_{13}(-\sqrt{\frac{2}{3}}D_5E_1-\frac{D_1E_2}{\sqrt2}+\frac{D_5E_2}{\sqrt6}+\frac{D_1E_3}{\sqrt2}+\frac{D_5E_3}{\sqrt6})+\\
r_{14}(D_2S_{\{1,2\}}+D_3S_{\{1,3\}}+D_3S_{\{2,3\}})+\\
r_{15}(-D_2A_{\{1,2\}}+D_3A_{{1,3}}-D_4A_{\{2,3\}})
\end{aligned}
\end{array}
\end{bmatrix},
\end{aligned}
\end{equation}

Hence  the $2$-order $k\cdot p$ Hamiltonian, namely, $\mathcal{H}_2$ is the summation of $\ref{example-1-Hkp-3}$ and $\ref{example-1-Hkp-4}$ as in Eq. \ref{expand-l}.\\

\subsection{Imposing antiunitary symmetry}

When $G(\mathbf{k}^*)>H(\mathbf{k}*)$, namely, $G(\mathbf{k}^*)=H(\mathbf{k}^*)+A\cdot H(\mathbf{k}^*)$, first, $\xi_n$ may be in the form of $\{\alpha\}$, $\{\beta,\beta\}$ or $\gamma,\delta$. As in Eq. \ref{get-l}, the elementary $k\cdot p$ blocks in the form of $\{H^l_{\alpha\beta m}\}$ could be used to construct $\mathcal{H}^l_{nn'}$ which may be a block matrix. However, $A$ could impose more constraints so that these elementary blocks may be modified as described in the following: It may relate  the blocks of  $\mathcal{H}^l_{nn'}$ or it directly impose constraints on one block, thus resulting in new elementary $k\cdot p$ blocks denoted by $\bar{H}$.   Firstly, consider the diagonal blocks, namely, $n=n'$. In this case, $\xi_n=\xi_{n'}$. We distinguish between two cases, cases 1.1 and 1.2 as follows.

\begin{itemize}
  \item \textbf{Case 1.1}: $\xi_n=\{\alpha\}$. $\mathcal{H}^l_{n,n}(\mathbf{q})$ can be written by $\sum_m r_m H^l_{\alpha\alpha m}(\mathbf{q})$ satisfying Eq. \ref{symm-constraints-1} where $r_m$ is a real parameter, namely, $r_m\in\mathbb{R}$,  so that $\mathcal{H}^l_{n,n}(\mathbf{q})$ is Hermitian.
  In this case, $D^{\xi_n}(A)=u_\alpha \mathcal{K}$, thus, from Eq. \ref{symm-constraints-2}, $\{r_m\}$ cannot be freely chosen in general and they should satisfy:
\begin{equation}\label{case1.1}
    \sum_{m'}\mathcal{U}_{mm'}r_{m'}=r_m,
\end{equation}
where $\mathcal{U}$ is a real matrix and can be obtained by $u_{\alpha}{H^l_{\alpha\alpha m}(-A_0^{-1}\mathbf{q})}^* u_{\alpha}^\dag =\sum_{m'} \mathcal{U}_{m'm} H^l_{\alpha\alpha m'}(\mathbf{q})$ where $A_0$ is the point part of $A$, for example, when $A=\{\beta_a|\boldsymbol{\tau}_a\}\Theta$, $A_0=\beta_a$. Denote the solution of Eq. \ref{case1.1} as: $r_m=\sum_{\bar{m}}r_{\bar{m}} b^{\bar{m}}_m$, where $\bar{m}$ labels the basis vectors $b^{\bar{m}}$ of the solution space of Eq. \ref{case1.1}. Thus $\mathcal{H}^{l}_{n,n}$ satisfying Eqs. \ref{symm-constraints-1} and \ref{symm-constraints-2} is:
\begin{equation}\label{case1.1-2}
    \mathcal{H}^l_{n,n}(\mathbf{q})=\sum_{\bar{m}} r_{\bar{m}} \bar{H}^l_{\alpha\alpha\bar{m}}(\mathbf{q}),
\end{equation}
where $\bar{H}^l_{\alpha\alpha \bar{m}}(\mathbf{q})=\sum_{m} b^{\bar{m}}_m H^l_{\alpha\alpha m}(\mathbf{q})$ and $r_{\bar{m}}\in \mathbb{R}$.

  \item \textbf{Case 1.2}: $\xi_n=\{\alpha,\beta\}$.  Based on Eq. \ref{get-l}, $\mathcal{H}^l_{n,n}$ can be written as,
 \begin{equation}\label{case1.2-0}
    \mathcal{H}^{l}_{n,n}(\mathbf{q})=\left(\begin{array}{cc}
 W_{11}(\mathbf{q})& W_{12}(\mathbf{q})\\
 W_{21}(\mathbf{q})& W_{22}(\mathbf{q})
\end{array}\right),
 \end{equation}
where $W_{21}(\mathbf{q})=W_{12}(\mathbf{q})^\dag$ to guarantee the Hermiticity.

In this case, $D^{\xi_n}(A)=\left(\begin{array}{cc}0& u_\beta\\u_\alpha& 0 \end{array}\right) \mathcal{K}$: When $\alpha=\beta$, $u_\beta=u'_\alpha$. Thus, from Eq. \ref{symm-constraints-2},
\begin{equation}\label{case1.2-1}
 W_{22}(\mathbf{q})=u_\alpha W_{11}(-A_0^{-1}\mathbf{q})^* u_\alpha^\dag,
\end{equation},
so that given $W_{11}$ we would know $W_{22}$, and $W_{11}$ is given by
\begin{equation}\label{case1.2-2}
    W_{11}(\mathbf{q})=\sum_{m}r_{m} H^l_{\alpha\alpha m}(\mathbf{q}), r_m \in \mathbb{R}.
\end{equation}

From Eq. \ref{symm-constraints-2} combined with $W_{12}=W_{21}^\dag$, we also have,
\begin{equation}\label{case1.2-3}
    u_\beta W_{12}(-A_0^{-1}\mathbf{q})^\top u_{\alpha}^\dag=W_{12}(\mathbf{q}),
\end{equation}
and writing $W_{12}(\mathbf{q})=\sum_{m}c_m H^l_{\alpha\beta m}(\mathbf{q})$ where $c_m$ is a complex parameter, namely, $c_m\in\mathbb{C}$, so that it satisfies Eq. \ref{symm-constraints-1} naturally, while Eq. \ref{case1.2-3} would impose additional constraints on $\{c_m\}$ as follows,
\begin{equation}\label{case1.2-4}
    \sum_{m'}\mathcal{U}_{mm'}c_{m'}=c_m,
\end{equation}
where $\mathcal{U}$ can be found by $u_{\beta}{H^l_{\alpha\beta m}(-A_0^{-1}\mathbf{q})^\top} u_{\alpha}^\dag =\sum_{m'} \mathcal{U}_{m'm} H^l_{\alpha\beta m'}(\mathbf{q})$. Denoting the solution of Eq. \ref{case1.2-4} by $c_{m}=\sum_{\bar{m}}c_{\bar{m}}b^{\bar{m}}_m$,
\begin{equation}\label{case1.2-5}
    W_{12}(\mathbf{q})=\sum_{\bar{m}}c_{\bar{m}}\bar{H}_{\alpha\beta \bar{m}},
\end{equation}
would satisfy Eqs. \ref{symm-constraints-1} and \ref{symm-constraints-2} at the same time. In Eq. \ref{case1.2-5}, $\bar{H}^l_{\alpha\beta m}(\mathbf{q})=\sum_{m}b^{\bar{m}}_m H^l_{\alpha\beta m}(\mathbf{q})$ and $c_{\bar{m}}\in\mathbb{C}$.
\end{itemize}

Next, we consider the nondiagonal blocks $n\ne n'$. We would distinguish between the following five cases:
\begin{itemize}
  \item \textbf{2.1}:   $\xi_n=\{\alpha\}$ and $\xi_{n'}=\{\alpha\}$. Firstly, $\mathcal{H}^l_{nn'}(\mathbf{q})$ can be written as,
\[\mathcal{H}^l_{nn'}(\mathbf{q})=\sum_m c_m H^l_{\alpha\alpha m}(\mathbf{q}), c_m\in\mathbb{C},\]
which satisfies Eq. \ref{symm-constraints-1}. Then consider Eq. \ref{symm-constraints-2}. We then find that $\{c_m\}$ should satisfy,
\begin{equation}\label{case2.1-1}
   \sum_{m'}\mathcal{U}_{mm'}c_{m'}^*=c_m,
\end{equation}
where $\mathcal{U}$ is real and can be obtained by $u_{\alpha}{H^l_{\alpha\alpha m}(-A_0^{-1}\mathbf{q})}^* u_{\alpha}^\dag =\sum_{m'} \mathcal{U}_{m'm} H^l_{\alpha\alpha m'}(\mathbf{q})$ as case 1.1.  In order to find solutions to Eq. \ref{case2.1-1}, we can first written $c_m=r^1_m+\mathrm{i}r^2_m$ where $r^l_m,r^2_m\in\mathbb{R}$, and then Eq. \ref{case2.1-1} is changed to:
\begin{equation}\label{case2.1-2-1}
\sum_{m'}\mathcal{U}_{mm'}r^1_{m'}=r^1_m,
\end{equation}
and
\begin{equation}\label{case2.1-2-2}
-\sum_{m'}\mathcal{U}_{mm'}r^2_{m'}=r^2_m.
\end{equation}
Eq. \ref{case2.1-2-1} is the same as Eq. \ref{case1.1}.  Denote the solutions as $r^1_m=\sum_{\bar{m}} r_{\bar{m}}  b^{\bar{m}}_m$ and $r^2_m=\sum_{\bar{m}''} r''_{\bar{m}''}  {b''}^{\bar{m}''}_m$, respectively, where $r_{\bar{m}},r''_{\bar{m}''}\in\mathbb{R}$. We then have,
\begin{equation}\label{case2.1-3}
    \mathcal{H}^l_{nn'}(\mathbf{q})=\sum_{\bar{m}'}{r'}_{\bar{m}'} {\bar{H'}}^l_{\alpha\alpha\bar{m}'}(\mathbf{q}),
\end{equation}
which satisfies Eqs. \ref{symm-constraints-1} and \ref{symm-constraints-2}. In Eq. \ref{case2.1-3}, ${\bar{H'}}^l_{\alpha\alpha\bar{m}'}(\mathbf{q})=\sum_{m}b^{\bar{m}}_m H^l_{\alpha\alpha m}(\mathbf{q})$ or  ${\bar{H'}}^l_{\alpha\alpha \bar{m}'}(\mathbf{q})=\sum_{m}b^{\bar{m}''}_m H^l_{\alpha\alpha m}(\mathbf{q})$, and $r'_{\bar{m}'}\in\mathbb{R}$. 

  \item \textbf{2.2}: $\xi_n=\{\alpha\},\xi_{n'}=\{\beta\}$ and $\alpha\ne\beta$. In this case, as before, we first write $\mathcal{H}^l_{nn'}(\mathbf{q})=\sum_{m}c_mH^l_{\alpha\beta m}(\mathbf{q})$ where $c_m\in\mathbb{C}$, which satisfies Eq. \ref{symm-constraints-1}. Then we consider the effect of $A$. From Eq. \ref{symm-constraints-2}, we have,
\begin{equation}\label{case2.2}
    \sum_{m'}\mathcal{U}_{mm'}c_{m'}^*=c_m.
\end{equation}
where $\mathcal{U}$  can be obtained by $u_{\alpha}{H^l_{\alpha\beta m}(-A_0^{-1}\mathbf{q})}^* u_{\beta}^\dag =\sum_{m'} \mathcal{U}_{m'm} H^l_{\alpha\beta m'}(\mathbf{q})$ and can be a complex matrix. To find the solutions $\{c_m\}$ of Eq. \ref{case2.2}, we first write $c_m=r^1_m+\mathrm{i}r^2_m$ where $r^1_m,r^2_m\in\mathbb{R}$ as well as $\mathcal{U}=\mathcal{U}^1+\mathrm{i}\mathcal{U}^2$ where $\mathcal{U}^1$ and $\mathcal{U}^2$ are real matrices. Then Eq. \ref{case2.2} changes to,
\begin{equation}\label{2.2-1}
\left(\begin{array}{cc}
\mathcal{U}^1& \mathcal{U}^2\\
\mathcal{U}^2& -\mathcal{U}^1
\end{array}\right)\left(\begin{array}{c}r^1\\r^2\end{array}\right)=\left(\begin{array}{c}r^1\\r^2\end{array}\right),
\end{equation}
where $r^1$ and $r^2$ are column vectors composed of $r^1_m$ and $r^2_m$. Suppose the solutions of Eq. \ref{2.2-1} are $\{b^{\bar{m}}\}$ which are all real and $b^{\bar{m}}=\left(\begin{array}{c} {b^1}^{\bar{m}}\\{b^2}^{\bar{m}}\end{array} \right)$, namely splitting it in half. $c_m$ is thus equal to $\sum_{\bar{m}}r_{\bar{m}}({b^1}^{\bar{m}}_m+\mathrm{i}{b^2}^{\bar{m}}_m), r_{\bar{m}}\in\mathbb{R}$, and we will have,
\begin{equation}\label{2.2-2}
\mathcal{H}^l_{nn'}(\mathbf{q})=\sum_{\bar{m}}r_{\bar{m}}\bar{H}^l_{\alpha\beta\bar{m}}(\mathbf{q}),
\end{equation}
where $r_{\bar{m}}\in\mathbb{R}$ and $\bar{H}^l_{\alpha\beta\bar{m}}(\mathbf{q})=\sum_{m}({b^1}^{\bar{m}}_m+\mathrm{i}{b^2}^{\bar{m}}_m)H^l_{\alpha\beta m}(\mathbf{q})$.
  \item \textbf{2.3}: $\xi_n=\{\alpha\}$ and $\xi_{n'}=\{\beta,\gamma\}$. In this case, we write $\mathcal{H}^l_{nn'}(\mathbf{q})$ as:
 \begin{equation}\label{case2.3}
    \mathcal{H}^l_{nn'}(\mathbf{q})=\left(\begin{array}{cc}W_{11}(\mathbf{q})& W_{12}(\mathbf{q})\end{array}\right),
 \end{equation}
where $W_{11}(\mathbf{q})=\sum_{m}c_m H^l_{\alpha\beta m}(\mathbf{q}),c_m\in\mathbb{C}$ satisfies Eq. \ref{symm-constraints}.  $W_{12}$ is related with $W_{11}$ through:
\begin{equation}\label{case2.3-1}
 W_{12}(\mathbf{q})=u_{\alpha}W_{11}(-A_0^{-1}\mathbf{q})^*u_{\beta}^\dag.
\end{equation}
Note that $W_{12}$ also satisfies Eq. \ref{symm-constraints-1}.
  \item \textbf{2.4}: $\xi_{n}=\{\alpha,\beta\}$ and $\xi_{n'}=\{\gamma\}$.
In this case, we write $\mathcal{H}^l_{nn'}(\mathbf{q})$ as:
 \begin{equation}\label{case2.4}
    \mathcal{H}^l_{nn'}(\mathbf{q})=\left(\begin{array}{c}W_{11}(\mathbf{q})\\ W_{21}(\mathbf{q})\end{array}\right),
 \end{equation}
where $W_{11}(\mathbf{q})=\sum_{m}c_m H^l_{\alpha\gamma m}(\mathbf{q}),c_m\in\mathbb{C}$ satisfies Eq. \ref{symm-constraints-1}.  $W_{21}$ is related with $W_{11}$ through:
\begin{equation}\label{case2.4-1}
 W_{21}(\mathbf{q})=u_{\alpha}W_{11}(-A_0^{-1}\mathbf{q})^*u_{\gamma}^\dag.
\end{equation}
  \item \textbf{2.5}: $\xi_{n}=\{\alpha,\beta\}$ and $\xi_{n'}=\{\gamma,\delta\}$.  Write $\mathcal{H}^l_{nn'}$ as:
\begin{equation}\label{2.5}
    \mathcal{H}^l_{nn'}(\mathbf{q})=\left(\begin{array}{cc}
W_{11}(\mathbf{q})& W_{12}(\mathbf{q})\\
W_{21}(\mathbf{q})& W_{22}(\mathbf{q})
\end{array}\right),
\end{equation}
where $W_{11}(\mathbf{q})=\sum_{m}c_m H^l_{\alpha\gamma m}(\mathbf{q})$ and $W_{12}(\mathbf{q})=\sum_{m'}c'_{m'} H^l_{\alpha\delta m'}(\mathbf{q})$ ($c_m,c'_m\in\mathbb{C}$). $W_{11}$ and $W_{12}$ naturally satisfy Eq. \ref{symm-constraints-1}. $A$ relates $W_{11}$ with $W_{22}$, $W_{12}$ with $W_{21}$, respectively:
\begin{equation}\label{case2.5-1}
   \begin{aligned}
& W_{22}(\mathbf{q})=u_{\alpha}W_{11}(-A_0^{-1}\mathbf{q})^*u_{\gamma}^\dag,\\
& W_{21}(\mathbf{q})=u_{\alpha}W_{12}(-A_0^{-1}\mathbf{q})^*u_{\delta}^\dag.
   \end{aligned}
\end{equation}

And when $\alpha=\beta$ or $\gamma=\delta$,  $\{H^l_{\alpha\gamma m}(\mathbf{q})\}$ are enough, and if $\alpha=\beta$,  we can obtain $W_{11}$ and $W_{21}$ firstly, then obtain $W_{22}$ and $W_{12}$ through $A$ while if $\gamma=\delta$, we can obtain $W_{11}$ and $W_{12}$ firstly, then obtain $W_{22}$ and $W_{21}$ through $A$. When $\alpha=\beta$ or $\gamma=\delta$ is not satisfied, we should know $\{H^{l}_{\alpha\gamma m}(\mathbf{q})\}$ and $\{H^{l}_{\alpha\delta m}(\mathbf{q})\}$. 
They are used to construct $W_{11}$ and $W_{12}$ firstly, and then we could obtain $W_{22}$ and $W_{21}$ through $A$.
 \end{itemize}

The above discussions can be summarized in the following two tables:
 \begin{table*}[!thbp]
\begin{tabular}{|c|c|c|c|c|c|c|}
  \hline
 \backslashbox{$\xi_n$}{$\xi_{n'}$} &$\{\alpha\}$& $\{\beta,\beta\}$& $\{\gamma,\delta\}$& $\{\alpha'\}$& $\{\beta',\beta'\}$&$\{\gamma',\delta'\}$\\\hline
  $\{\alpha\}$& {\color{purple}$\bar{H}^l_{\alpha\alpha\bar{m}}$},{\color{blue}$\bar{H'}^l_{\alpha\alpha\bar{m}'}$}& $H^l_{\alpha\beta m}$ &$H^l_{\alpha\gamma m}$& ${\color{blue}\bar{H}^l_{\alpha\alpha' \bar{m}}}$ &  $H^l_{\alpha\beta'm}$&  $H^l_{\alpha\gamma'm}$   \\\hline
  $\{\beta,\beta\}$& $H^l_{\beta\alpha m}$& $H^l_{\beta\beta m}$,{\color{blue}$\bar{H}^l_{\beta\beta\bar{m}}$} & $H^l_{\beta\gamma m}$ & $H^l_{\beta\alpha'm}$ & $H^l_{\beta\beta'm}$ & $H^l_{\beta\gamma'm}$  \\\hline
    $\{\gamma,\delta\}$& $H^l_{\gamma\alpha m}$& $H^l_{\gamma\beta m}$  &  $H^l_{\gamma\gamma m},{\color{orange}H^l_{\gamma\delta m}},{\color{blue}\bar{H}^l_{\gamma\delta\bar{m}}}$   & $H^l_{\gamma\alpha' m}$ & $H^l_{\gamma\beta'm}$ & $H^l_{\gamma\gamma'm},{\color{orange}H^l_{\gamma\delta'm}}$  \\\hline
  $\{\alpha'\}$&{\color{blue}$\bar{H}^l_{\alpha'\alpha\bar{m}}$} & $H^l_{\alpha'\beta m}$  & $H^l_{\alpha'\gamma m}$ & {\color{purple}$\bar{H}^l_{\alpha'\alpha'\bar{m}}$},{\color{blue}$\bar{H'}^l_{\alpha'\alpha'\bar{m}'}$} & $H^l_{\alpha'\beta'm}$ & $H^l_{\alpha'\gamma' m}$  \\\hline
  $\{\beta',\beta'\}$& $H^l_{\beta'\alpha m}$ & $H^l_{\beta'\beta m}$  & $H^l_{\beta'\gamma'}$ & $H^l_{\beta'\alpha'm}$ & $H^l_{\beta'\beta' m}$,{\color{blue}$\bar{H}^l_{\beta'\beta'\bar{m}}$} & $H^l_{\beta'\gamma'm}$  \\\hline
  $\{\gamma',\delta'\}$& $H^l_{\gamma'\alpha m}$& $H^l_{\gamma'\beta}$  & $H^l_{\gamma'\gamma m},{\color{orange}H^l_{\gamma'\delta'm}}$ & $H^l_{\gamma'\alpha'm}$ & $H^l_{\gamma'\beta'm}$ & $H^l_{\gamma'\gamma'm},{\color{orange}H^l_{\gamma'\delta' m}},{\color{blue}\bar{H}^l_{\gamma'\delta'\bar{m}}}$  \\
  \hline
\end{tabular}\caption{The elementary $k\cdot p$ blocks for $G(\mathbf{k}^*)>H(\mathbf{k}^*)$, namely, the little group of $\mathbf{k}^*$ contains  at least one antiunitary symmetry, denoted by $A$. In this case, the energy levels at $\mathbf{k}^*$ can be attributed by the co-irreps of $G(\mathbf{k}^*)$, $\xi_{n}=\{\alpha\},\{\beta,\beta\}$ or $\{\gamma,\delta\}$. Here we use different symbols to denote different irreps of $H(\mathbf{k}^*)$. The constructions of $\mathcal{H}^l_{nn'}$ for all these situations are listed in Table \ref{AA} below. The explicit results for a specific type of elementary $k\cdot p$ blocks can be found in SM: Part III \cite{SM}, printed in the same color shown in this table.}\label{A}
\end{table*}

\begin{table*}[!tbhp]\tiny
\begin{tabular}{|c|c|c|}
\hline
$\begin{aligned}&n=n',\xi_n=\xi_{n'}=\{\alpha\}: \\ &\mathcal{H}^l_{nn'}=r_{\bar{m}}\bar{H}^l_{\alpha\alpha\bar{m}}\end{aligned}$ &

$\begin{aligned}&n=n',\xi_n=\xi_{n'}=\{\beta,\beta\}: \mathcal{H}^l_{nn'}=\\&\left(\begin{array}{cc}r_m H^l_{\beta\beta m}& c_{\bar{m}}\bar{H}^l_{\beta\beta\bar{m}}\\
h.c.&  r_m u_\beta {H^l_{\beta\beta m}(-A_0^{-1}\mathbf{q})}^*u_\beta^\dag
\end{array}\right)\end{aligned}$&

$\begin{aligned}&n=n', \xi_{n}=\xi_{n'}=\{\gamma,\delta\}: \mathcal{H}^l_{nn'}=\\&\left(\begin{array}{cc} r_mH^l_{\gamma\gamma m}&  c_{\bar{m}}\bar{H}^l_{\gamma\delta\bar{m}}\\
h.c.&  r_m u_\gamma {H^l_{\gamma\gamma m}(-A_0^{-1}\mathbf{q})}^*u_\gamma^\dag\end{array}\right)\end{aligned}$\\
\hline\hline

$\begin{aligned}&n\ne n', \xi_n=\xi_{n'}=\{\alpha\}: \\  &\mathcal{H}^l_{nn'}=r_{\bar{m}'}\bar{H'}^l_{\alpha\alpha\bar{m}'}\end{aligned}$&
$\begin{aligned}&n\ne n', \xi_n=\xi_{n'}=\{\beta,\beta\}: \mathcal{H}^l_{nn'}=\\&\left(\begin{array}{cc}c_m H^l_{\beta\beta m}& c'_{m}H_{\beta\beta m}\\
  {c'_{m}}^*u_{\beta}H^l_{\beta\beta m}(-A_0^{-1}\mathbf{q})^*{u'_\beta}^\dag&  c_m^* u_\beta {H^l_{\beta\beta m}(-A_0^{-1}\mathbf{q})}^*u_\beta^\dag
\end{array}\right)\end{aligned}$&

$\begin{aligned}&n\ne n', \xi_{n}=\xi_{n'}=\{\gamma,\delta\}: \mathcal{H}^l_{nn'}=\\&\left(\begin{array}{cc}c_mH^l_{\gamma\gamma m}& c'_{m}H^l_{\gamma\delta m}\\
{c'_{m}}^*u_{\gamma}H^l_{\gamma\delta m}(-A_0^{-1}\mathbf{q})^*u_\delta^\dag& c^*_m u_\gamma {H^l_{\gamma\gamma m}(-A_0^{-1}\mathbf{q})}^*u_\gamma^\dag\end{array}\right)\end{aligned}$\\
\hline

$\begin{aligned}&n\ne n', \xi_n=\{\alpha\},\xi_{n'}=\{\alpha'\}: \\ &\mathcal{H}^l_{nn'}=c_m \bar{H}^l_{\alpha\alpha'\bar{m}}\end{aligned}$&

$\begin{aligned}&n\ne n', \xi_n=\{\alpha\},\xi_{n'}=\{\beta,\beta\}: \mathcal{H}^l_{nn'}=\\&\left(\begin{array}{cc}c_mH^l_{\alpha\beta m}& c^*_mu_\alpha H^l_{\alpha\beta m}(-A_0^{-1}\mathbf{q})^*u_\beta^\dag\end{array}\right)\end{aligned}$&

$\begin{aligned}&n\ne n', \xi_{n}=\{\alpha\},\xi_{n'}=\{\gamma,\delta\}: \mathcal{H}^l_{nn'}=\\&\left(\begin{array}{cc}c_mH^l_{\alpha\gamma m}&  c^*_mu_\alpha H^l_{\alpha\gamma m}(-A_0^{-1}\mathbf{q})^*u_\gamma^\dag\end{array}\right)\end{aligned}$
\\\hline

$\begin{aligned}&n\ne n', \xi_n=\{\beta,\beta\},\xi_{n'}=\{\alpha\}: \\ &\mathcal{H}^l_{nn'}=\left(\begin{array}{c}c_m H^l_{\beta\alpha m}\\  c_m^* u_\beta H^l_{\beta\alpha m}(-A_0^{-1}\mathbf{q})^* u_{\alpha}^\dag\end{array}\right)\end{aligned}$&

$\begin{aligned}& n\ne n', \xi_n=\{\beta,\beta\},\xi_{n'}=\{\gamma,\delta\}: \\ &\mathcal{H}^l_{nn'}=\left(\begin{array}{cc}c_mH^l_{\beta\gamma m}& {c'}^*_mu'_\beta H^l_{\beta\gamma m}(-A_0^{-1}\mathbf{q})^*u_{\gamma}^\dag\\
c'_mH^l_{\beta\gamma m}& c^*_mu_\beta H^l_{\beta\gamma m}(-A_0^{-1}\mathbf{q})^*u_{\gamma}^\dag\end{array}\right)\end{aligned}$&

$\begin{aligned}& n\ne n', \xi_n=\{\beta,\beta\},\xi_{n'}=\{\beta',\beta'\}: \mathcal{H}^l_{nn'}=\\&\left(\begin{array}{cc}c_mH^l_{\beta\beta' m}& {c'_m}^*u'_\beta H^l_{\beta\beta' m}(-A_0^{-1}\mathbf{q})^*{u'_{\beta'}}^\dag\\
c'_mH^l_{\beta\beta' m}& c_mH^l_{\beta\gamma m}\end{array}\right)\end{aligned}$\\\hline

$\begin{aligned}&n\ne n', \xi_n=\{\gamma,\delta\},\xi_{n'}=\{\alpha\}: \mathcal{H}^l_{nn'}=\\&\left(\begin{array}{c}  c_mH^l_{\gamma\alpha m}\\  c_m^*u_\gamma H^l_{\gamma\alpha}(-A_0^{-1}\mathbf{q})^*u_\alpha^\dag\end{array}\right)\end{aligned}$&

$\begin{aligned}&n\ne n', \xi_n=\{\gamma,\delta\},\xi_{n'}=\{\beta,\beta\}: \mathcal{H}^l_{nn'}=\\& \left(\begin{array}{cc}c_mH^l_{\gamma\beta m}&c'_mH^l_{\gamma\beta m}\\
{c'_m}^*u_{\gamma}H^l_{\gamma\beta m}(-A_0^{-1}\mathbf{q})^*{u'_{\beta}}^\dag& c^*_mu_\beta H^l_{\gamma\beta m}(-A_0^{-1}\mathbf{q})^*u_{\beta}^\dag
\end{array}\right)\end{aligned}$&

$\begin{aligned}&n\ne n', \xi_n=\{\gamma,\delta\},\xi_{n'}=\{\gamma',\delta'\}: \mathcal{H}^l_{nn'}=\\&\left(\begin{array}{cc}c_mH^l_{\gamma\gamma'm}&c'_mH^l_{\gamma\delta'm}\\
{c'_m}^*u_{\gamma}H^l_{\gamma\delta'm}(-A_0^{-1}\mathbf{q})^*u_{\delta'}^\dag& c_m^*u_\gamma H^l_{\gamma\gamma'm}(-A_0^{-1}\mathbf{q})^*u_{\gamma'}^\dag
\end{array}\right)\end{aligned}$\\\hline
\end{tabular}\caption{Construction of $\mathcal{H}^l_{nn'}$ when there is antiunitary symmetry. We list all possible situations: The antiunitary symmetry may result in three kinds of co-irreps in the form of $\{\alpha\}$, $\{\beta,\beta\}$ or $\{\gamma,\delta\}$ ($\gamma\ne\delta$ and we can always require that $\gamma<\delta$).  Hence, for diagonal blocks $n=n'$, there are three cases corresponding to three cases in Eq. \ref{app:herring} of the Appendix, while for nondiagonal blocks $n\ne n'$, there are 12  cases:  for  $\xi_n\ne\xi_{n'}$, there are $3^2$ cases while for $\xi_n=\xi_{n'}$ there are 3 cases.      These formulas are consistent with  Table \ref{A}. For example, $\xi_n=\xi_{n'}=\{\alpha\}$ appears  in two cases  $n=n'$ and $n\ne n'$, for which the construction of $\mathcal{H}^l_{nn'}$ needs $\bar{H}^l_{\alpha\alpha \bar{m}}$ and $\bar{H'}^l_{\alpha\alpha\bar{m}'}$, respectively, both shown in Table \ref{A}.  $A_0$ and matrices like $u_{\alpha}$ are also given in SM: Part I \cite{SM} (and their definition can be found in  Sec. \ref{app:Aonirreps}). The Einstein  summation convention over $m, \bar{m}, \bar{m}'$ is adopted. }\label{AA}
\end{table*}

\newpage

\section{Exapmles}\label{app:detail}

\subsection{$k\cdot p$ models in graphene}\label{gra}
In this part, we use graphene which crystallizes in space group 191 to show that using our strategy and the elementary $k\cdot p$ blocks, we can obtain the low energy models around $K$ point for graphene very conveniently. Both models  when considering or neglecting spin-orbit coupling (SOC) are derived.   The MSG for space group 191 with TRS is MSG 191. 234 (type II).  As shown in Fig. \ref{kp-valley}, $K'$ is related with $K$ by inversion symmetry $I$ so that we can only consider $K$. For $K$ point, focusing on the energy window near the Fermi level shown in Fig. \ref{kp-valley}, we consider one energy level $E_a$ when SOC is neglected while two energy levels $E_b,E_c$ when SOC is considered,  as shown by solid dots in Fig. \ref{kp-valley}(a) and (b), respectively. When SOC is neglected,  as shown in SM: Part I \cite{SM} for $K$ of MSG 191.234 (the position of results for the co-irreps can be located  by clicking on the corresponding  page number in the index),  it owns 6 single-valued co-irreps: $\xi=\{1\},\{2\},\{3\},\{4\},\{5\},\{6\}$ (note that antiunitary symmetry is present), and the considered energy level is found to correspond to $\{6\}$ (which is two-dimensional): $\xi_{a}=\{6\}$. Consider the first-order $k\cdot p$ Hamiltonian $\mathcal{H}_{L=1}=\mathcal{H}_{L=1,aa}=\mathcal{H}^{l=1}_{aa}$:  since only one energy level is considered,  the only block is just the whole Hamiltonian. Based on Table \ref{A}, two kinds of elementary blocks should be used: $\bar{H}^1_{6,6,\bar{m}}$ and $\bar{H'}^1_{6,6,\bar{m}'}$. Then based on Table \ref{AA}, we only need $\bar{H}^1_{6,6,\bar{m}}$ since $n=n'=a$ which are printed in purple in SM: Part III \cite{SM}.  It is then found that $\bar{m}$ can only take 1, namely, there is only one such kind of elementary $k\cdot p$ block: $\bar{H}^1_{6,6,1}=-P_1S_{\{1,2\}}+\frac{E_1P_2-E_2P_2}{\sqrt 2}$, thus $\mathcal{H}^1_{a,a}=r_1 H^1_{6,6,1}=\frac{r_1}{\sqrt 2}\left(\begin{array}{cc}q_y& -q_x\\-q_x& -q_y\end{array}\right)$ (from Table \ref{AA}). This is simply the two dimensional massless Dirac Hamiltonian.

The $k\cdot p$ model with  SOC considered  can also be constructed easily.   It is found that $K$ owns 3 two-dimensional double-valued co-irreps:  $\xi=\{1\},\{2\},\{3\}$. The  two energy levels considered are found to be: $\xi_b=\{1\}$ and $\xi_{c}=\{3\}$. The $k\cdot p$ Hamiltonian in the first order is $\mathcal{H}_{L=1}$ which can be divided into blocks $\mathcal{H}_{L=1,bb}$, $\mathcal{H}_{L=1,bc}$, $\mathcal{H}_{L=1,cb}=\mathcal{H}_{L=1,bc}^\dag$ and $\mathcal{H}_{L=1,cc}$. For $\mathcal{H}_{L=1,bb}$ and $\mathcal{H}_{L=1,cc}$, Eq. \ref{expand-l} indicates that $\mathcal{H}^1_{bb}$ and $\mathcal{H}^1_{cc}$ should be known. Then from Tables \ref{A} and \ref{AA}, elementary $k\cdot p$ blocks: $\bar{H}^1_{1,1,\bar{m}}$ and $\bar{H}^1_{3,3,\bar{m}}$ should be used. However, they are found to be both $\{\}$, thus $\mathcal{H}^1_{bb}=\mathcal{H}^1_{cc}=0$. Then consider $\mathcal{H}_{1,bc}=\mathcal{H}^1_{bc}$, which need the knowledge of $\bar{H}^1_{1,3,\bar{m}}$ from Table \ref{A}. From SM: Part III \cite{SM}, $\bar{m}$ can only take 1 and  $H^1_{bc}=r_1\bar{H}^1_{1,3,1}=e^{\mathrm{i}\frac{\pi}{3}}\left(\begin{array}{cc}0& r_1(q_y+\mathrm{i} q_x)\\ r_1 (q_y-\mathrm{i}q_x)& 0 \end{array}\right)$. Hence consider the zeroth order Hamiltonian $\mathrm{dia}=(m,m,-m,-m)$ where $m$ is a real parameter, the whole $k\cdot p$ Hamiltonian in this case to first order is  $\left(\begin{array}{cc}m& r_1(q_y+\mathrm{i} q_x)\\ h.c.& -m \end{array}\right)\oplus \left(\begin{array}{cc}m& r_1(q_y-\mathrm{i} q_x)\\ h.c.& -m \end{array}\right)$ by rearranging the basis kets.

\begin{figure}
  \includegraphics[width=9cm]{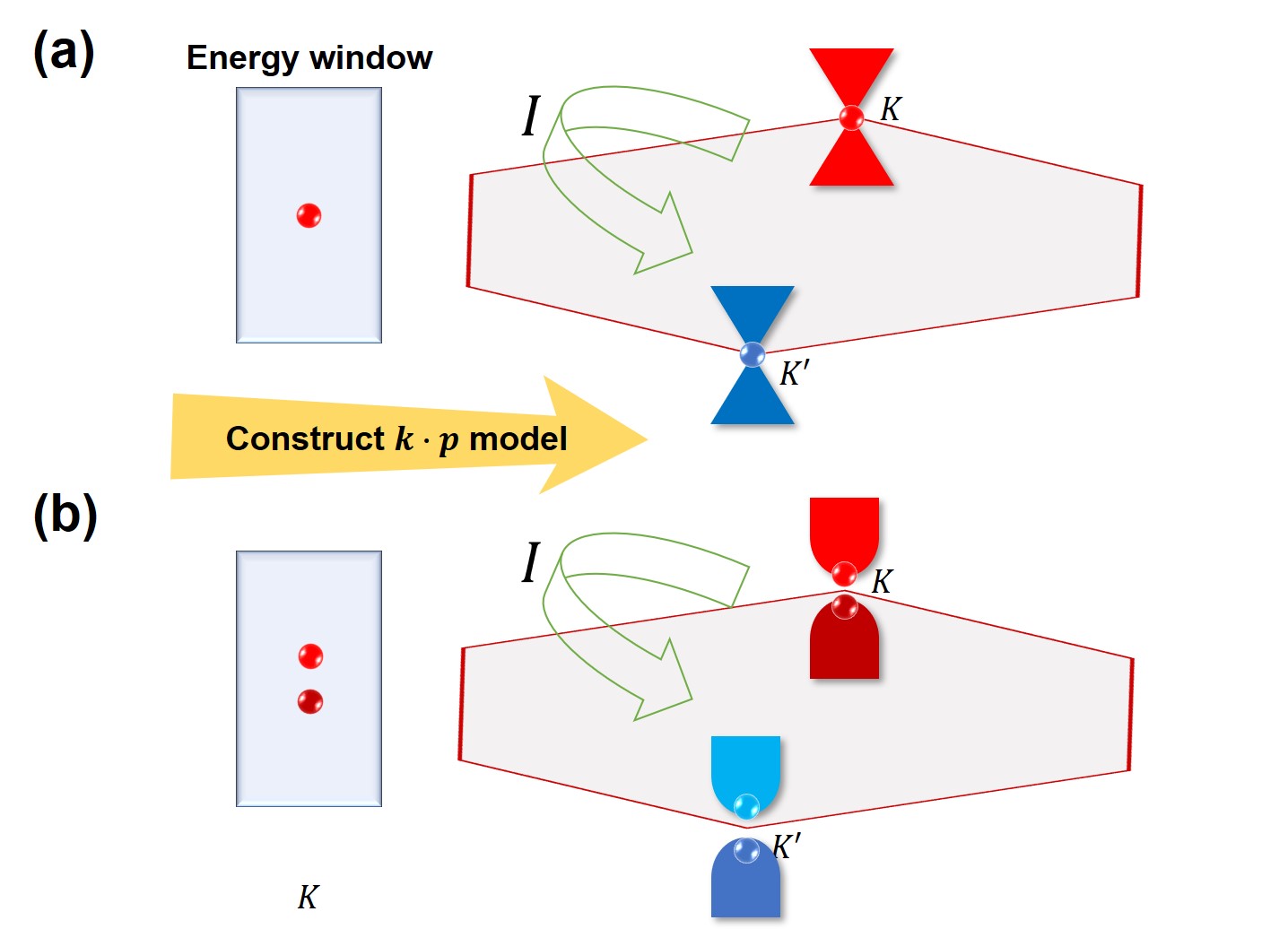}\\
  \caption{Demonstration of constructions of low-energy models for graphene: (a) SOC is neglected; (b) SOC is considered. }\label{kp-valley}
\end{figure}

\subsection{First-order eight-band Kane model}\label{app:detail-216}
\subsubsection{Deriving first-order  eight-band Kane model from our results}
For the derivation of the first-order eight-band Kane model for zincblende structures, firstly one need to identify the MSG, which is found to be   SG 216 or type- II MSG (MSG216.75). We consider the double-valued representations for $\Gamma$ point. As shown in SM: Part I \cite{SM},  it owns three co-irreps denoted as $\{1\},\{2\},\{3\}$ whose dimensions are $2,2,4$ respectively. Consider three energy levels $E_a=\epsilon_1$,$E_b=\epsilon_2$, $E_c=\epsilon_3$ whose co-irreps are $\xi_a=\{1\}$, $\xi_b=\{2\}$ and $\xi_c=\{3\}$. Hence we obtain an eight-band model. In the following, we calculate the first-order $k\cdot p$ Hamiltonian $\mathcal{H}_{L=1}(\mathbf{q})=\mathcal{H}^{l=1}(\mathbf{q})$. Note that the $0$-order Hamiltonian is simply equal to $\mathrm{dia}(\epsilon_1,\epsilon_1,\epsilon_2,\epsilon_2,\epsilon_3,\epsilon_3,\epsilon_3,\epsilon_3)$. For $\mathcal{H}^1_{aa}$, which belongs to case 1.1,  we should use elementary $k\cdot p$ blocks: $\{\bar{H}^1_{1,1,\bar{m}}\}$ (indicated by $\{1\}\&\{1\}$ in SM: Part III \cite{SM}) as from Table \ref{AA}, and as shown in SM: Part III \cite{SM}, it is  $\{\}$.
For $\mathcal{H}^1_{bb}$, we should use elementary $k\cdot p$ blocks for $\{2\}\&\{2\}$: $\{\bar{H}^1_{22,\bar{m}}\}$ and as shown in  SM: Part III \cite{SM}, it is also $\{\}$. For $\mathcal{H}^1_{cc}$, we should use elementary $k\cdot p$ blocks for $\{3\}\&\{3\}$: $\{\bar{H}^1_{3,3,\bar{m}}\}$ and as shown in SM: Part III \cite{SM}, it is: $\{-\frac{1}{2} \sqrt{\frac{3}{2}} P_2 A_{\{1,3\}}+\frac{1}{2} \sqrt{\frac{3}{2}} P_2 A_{\{2,4\}}-\frac{P_2 A_{\{1,4\}}}{2 \sqrt{2}}-\frac{P_2 A_{\{2,3\}}}{2 \sqrt{2}}+\frac{P_3 S_{\{1,2\}}}{\sqrt{2}}-\frac{1}{2} \sqrt{\frac{3}{2}} P_1 S_{\{1,3\}}+\frac{P_1 S_{\{1,4\}}}{2 \sqrt{2}}+\frac{P_1 S_{\{2,3\}}}{2 \sqrt{2}}+\frac{1}{2} \sqrt{\frac{3}{2}} P_1 S_{\{2,4\}}-\frac{P_3 S_{\{3,4\}}}{\sqrt{2}}\}$, so that, $\mathcal{H}^1_{cc}=r_1 (-\frac{1}{2} \sqrt{\frac{3}{2}} P_2 A_{\{1,3\}}+\frac{1}{2} \sqrt{\frac{3}{2}} P_2 A_{\{2,4\}}-\frac{P_2 A_{\{1,4\}}}{2 \sqrt{2}}-\frac{P_2 A_{\{2,3\}}}{2 \sqrt{2}}+\frac{P_3 S_{\{1,2\}}}{\sqrt{2}}-\frac{1}{2} \sqrt{\frac{3}{2}} P_1 S_{\{1,3\}}+\frac{P_1 S_{\{1,4\}}}{2 \sqrt{2}}+\frac{P_1 S_{\{2,3\}}}{2 \sqrt{2}}+\frac{1}{2} \sqrt{\frac{3}{2}} P_1 S_{\{2,4\}}-\frac{P_3 S_{\{3,4\}}}{\sqrt{2}})$.  Explicitly write $\mathcal{H}^1_{cc}(\mathbf{q})$ as:

\begin{equation}\label{examples-2-1}
r_1\left(
\begin{array}{cccc}
 0 & \frac{q_z}{2} & -\frac{\sqrt{3} q_x}{4}+\frac{1}{4} i \sqrt{3} q_y & \frac{q_x}{4}+\frac{i q_y}{4} \\
 \frac{q_z}{2} & 0 & \frac{q_x}{4}+\frac{i q_y}{4} & \frac{\sqrt{3} q_x}{4}-\frac{1}{4} i \sqrt{3} q_y \\
 -\frac{1}{4} \sqrt{3} q_x-\frac{1}{4} i \sqrt{3} q_y & \frac{q_x}{4}-\frac{i q_y}{4} & 0 & -\frac{q_z}{2} \\
 \frac{q_x}{4}-\frac{i q_y}{4} & \frac{\sqrt{3} q_x}{4}+\frac{1}{4} i \sqrt{3} q_y & -\frac{q_z}{2} & 0 \\
\end{array}
\right).
\end{equation}

For $\mathcal{H}^{1}_{ab}$, we should use elementary $k\cdot p$ blocks for $\{1\}\&\{2\}$: $\{\bar{H}^1_{1,2,\bar{m}}\}$, and from  SM: Part III \cite{SM}, it is,
$\{-\frac{P_1 M_{\{1,2\}}}{\sqrt{2}}-\frac{P_1 M_{\{2,1\}}}{\sqrt{2}}+\frac{\mathrm{i} P_2 M_{\{2,1\}}}{\sqrt{2}}+\frac{P_3 M_{\{1,1\}}}{\sqrt{2}}-\frac{\mathrm{i} P_2 M_{\{1,2\}}}{\sqrt{2}}-\frac{P_3 M_{\{2,2\}}}{\sqrt{2}}\}$, thus,
$\mathcal{H}^1_{ab}=r_2(-\frac{P_1 M_{\{1,2\}}}{\sqrt{2}}-\frac{P_1 M_{\{2,1\}}}{\sqrt{2}}+\frac{\mathrm{i} P_2 M_{\{2,1\}}}{\sqrt{2}}+\frac{P_3 M_{\{1,1\}}}{\sqrt{2}}-\frac{\mathrm{i}P_2 M_{\{1,2\}}}{\sqrt{2}}-\frac{P_3 M_{\{2,2\}}}{\sqrt{2}})$.  Write $\mathcal{H}^1_{ab}$ explicitly as:

\begin{equation}\label{examples-2-2}
r_2\left(
\begin{array}{cc}
 \frac{q_z}{\sqrt{2}} & -\frac{q_x}{\sqrt{2}}-\frac{\mathrm{i}q_y}{\sqrt{2}} \\
 -\frac{q_x}{\sqrt{2}}+\frac{\mathrm{i} q_y}{\sqrt{2}} & -\frac{q_z}{\sqrt{2}} \\
\end{array}
\right).
\end{equation}

For $\mathcal{H}^{1}_{ac}$, we should use elementary $k\cdot p$ blocks for $\{1\}\&\{3\}$: $\{\bar{H}^1_{1,3,\bar{m}}\}$, and from  SM: Part III \cite{SM}, it is,
$\{\frac{1}{2} \sqrt{\frac{3}{2}} P_1 M_{\{1,3\}}+\frac{P_1 M_{\{1,4\}}}{2 \sqrt{2}}+\frac{1}{2} \sqrt{\frac{3}{2}} P_1 M_{\{2,1\}}+\frac{P_1 M_{\{2,2\}}}{2 \sqrt{2}}-\frac{1}{2} \mathrm{i}\sqrt{\frac{3}{2}} P_2 M_{\{1,3\}}+\frac{\mathrm{i} P_2 M_{\{1,4\}}}{2 \sqrt{2}}+\frac{1}{2} \mathrm{i} \sqrt{\frac{3}{2}} P_2 M_{\{2,1\}}+\frac{P_3 M_{\{1,2\}}}{\sqrt{2}}-\frac{P_3 M_{\{2,4\}}}{\sqrt{2}}-\frac{i P_2 M_{\{2,2\}}}{2 \sqrt{2}}\}$, thus,
$\mathcal{H}^1_{ac}=r_3(\frac{1}{2} \sqrt{\frac{3}{2}} P_1 M_{\{1,3\}}+\frac{P_1 M_{\{1,4\}}}{2 \sqrt{2}}+\frac{1}{2} \sqrt{\frac{3}{2}} P_1 M_{\{2,1\}}+\frac{P_1 M_{\{2,2\}}}{2 \sqrt{2}}-\frac{1}{2} \mathrm{i}\sqrt{\frac{3}{2}} P_2 M_{\{1,3\}}+\frac{\mathrm{i} P_2 M_{\{1,4\}}}{2 \sqrt{2}}+\frac{1}{2} \mathrm{i} \sqrt{\frac{3}{2}} P_2 M_{\{2,1\}}+\frac{P_3 M_{\{1,2\}}}{\sqrt{2}}-\frac{P_3 M_{\{2,4\}}}{\sqrt{2}}-\frac{i P_2 M_{\{2,2\}}}{2 \sqrt{2}})$. Write $\mathcal{H}^1_{ac}$ explicitly as:

\begin{equation}\label{examples-2-3}
r_3\left(
\begin{array}{cccc}
 0 & \frac{q_z}{\sqrt{2}} & \frac{1}{2} \sqrt{\frac{3}{2}} q_x-\frac{1}{2} \mathrm{i} \sqrt{\frac{3}{2}} q_y & \frac{q_x}{2 \sqrt{2}}+\frac{\mathrm{i}q_y}{2 \sqrt{2}} \\
 \frac{1}{2} \sqrt{\frac{3}{2}} q_x+\frac{1}{2} \mathrm{i} \sqrt{\frac{3}{2}} q_y & \frac{q_x}{2 \sqrt{2}}-\frac{\mathrm{i} q_y}{2 \sqrt{2}} & 0 & -\frac{q_z}{\sqrt{2}} \\
\end{array}
\right).
\end{equation}

For $\mathcal{H}^{1}_{bc}$, we should use elementary $k\cdot p$ blocks for $\{2\}\&\{3\}$: $\{\bar{H}^1_{2,3,\bar{m}}\}$, and from  SM: Part III \cite{SM}, it is,
$\{\frac{P_1 M_{\{1,3\}}}{2 \sqrt{2}}-\frac{1}{2} \sqrt{\frac{3}{2}} P_1 M_{\{1,4\}}+\frac{P_1 M_{\{2,1\}}}{2 \sqrt{2}}-\frac{1}{2} \sqrt{\frac{3}{2}} P_1 M_{\{2,2\}}+\frac{\mathrm{i} P_2 M_{\{1,3\}}}{2 \sqrt{2}}+\frac{1}{2} \mathrm{i} \sqrt{\frac{3}{2}} P_2 M_{\{1,4\}}-\frac{1}{2} \mathrm{i}\sqrt{\frac{3}{2}} P_2 M_{\{2,2\}}+\frac{P_3 M_{\{1,1\}}}{\sqrt{2}}-\frac{P_3 M_{\{2,3\}}}{\sqrt{2}}-\frac{i P_2 M_{\{2,1\}}}{2 \sqrt{2}}\}$,

 thus,
$\mathcal{H}^1_{bc}=r_4(\frac{P_1 M_{\{1,3\}}}{2 \sqrt{2}}-\frac{1}{2} \sqrt{\frac{3}{2}} P_1 M_{\{1,4\}}+\frac{P_1 M_{\{2,1\}}}{2 \sqrt{2}}-\frac{1}{2} \sqrt{\frac{3}{2}} P_1 M_{\{2,2\}}+\frac{\mathrm{i} P_2 M_{\{1,3\}}}{2 \sqrt{2}}+\frac{1}{2} \mathrm{i} \sqrt{\frac{3}{2}} P_2 M_{\{1,4\}}-\frac{1}{2} \mathrm{i} \sqrt{\frac{3}{2}} P_2 M_{\{2,2\}}+\frac{P_3 M_{\{1,1\}}}{\sqrt{2}}-\frac{P_3 M_{\{2,3\}}}{\sqrt{2}}-\frac{i P_2 M_{\{2,1\}}}{2 \sqrt{2}})$.

Write $\mathcal{H}^1_{bc}$ explicitly as:

\begin{equation}\label{examples-2-4}
r_4 \left(
\begin{array}{cccc}
 \frac{q_z}{\sqrt{2}} & 0 & \frac{q_x}{2 \sqrt{2}}+\frac{\mathrm{i} q_y}{2 \sqrt{2}} & -\frac{1}{2} \sqrt{\frac{3}{2}} q_x+\frac{1}{2} i \sqrt{\frac{3}{2}} q_y \\
 \frac{q_x}{2 \sqrt{2}}-\frac{\mathrm{i} q_y}{2 \sqrt{2}} & -\frac{1}{2} \sqrt{\frac{3}{2}} q_x-\frac{1}{2} i \sqrt{\frac{3}{2}} q_y & -\frac{q_z}{\sqrt{2}} & 0 \\
\end{array}
\right).
\end{equation}

Finally, the first-order $k\cdot p$ Hamiltonian should be:

\begin{equation}\label{examples-2}
\mathcal{H}_{216}(\mathbf{q})=\left(
\begin{array}{cccccccc}
 0 & 0 & r_2 q_z & -q_+ r_2 & 0 & r_3 q_z & \frac{1}{2} \sqrt{3} q_- r_3 & \frac{q_+ r_3}{2} \\
 0 & 0 & -q_- r_2 & r_2 \left(-q_z\right) & \frac{1}{2} \sqrt{3} q_+ r_3 & \frac{q_- r_3}{2} & 0 & r_3 \left(-q_z\right) \\
 r_2 q_z & -q_+ r_2 & 0 & 0 & r_4 q_z & 0 & \frac{q_+ r_4}{2} & -\frac{1}{2} \sqrt{3} q_- r_4 \\
 -q_- r_2 & r_2 \left(-q_z\right) & 0 & 0 & \frac{q_- r_4}{2} & -\frac{1}{2} \sqrt{3} q_+ r_4 & r_4 \left(-q_z\right) & 0 \\
 0 & \frac{1}{2} \sqrt{3} q_- r_3 & r_4 q_z & \frac{q_+ r_4}{2} & 0 & 2 r_1 q_z & -\sqrt{3} q_- r_1 & q_+ r_1 \\
 r_3 q_z & \frac{q_+ r_3}{2} & 0 & -\frac{1}{2} \sqrt{3} q_- r_4 & 2 r_1 q_z & 0 & q_+ r_1 & \sqrt{3} q_- r_1 \\
 \frac{1}{2} \sqrt{3} q_+ r_3 & 0 & \frac{q_- r_4}{2} & r_4 \left(-q_z\right) & -\sqrt{3} q_+ r_1 & q_- r_1 & 0 & -2 r_1 q_z \\
 \frac{q_- r_3}{2} & r_3 \left(-q_z\right) & -\frac{1}{2} \sqrt{3} q_+ r_4 & 0 & q_- r_1 & \sqrt{3} q_+ r_1 & -2 r_1 q_z & 0 \\
\end{array}
\right)
\end{equation}
where $q_{\pm}=q_x\pm\mathrm{i}q_y$,  $r_1$ has been multiplied by $4$ compared with Eq. \ref{examples-2-1} and $r_2,r_3,r_4$ have been multiplied by $\sqrt 2$ compared with Eqs. \ref{examples-2-2},\ref{examples-2-3},\ref{examples-2-4}, respectively.

\subsubsection{Transformation}
The basis vectors in deducing the original eight-band Kane model \cite{kpbook} are listed as follows:
\begin{equation}\label{app:kane-basis}
    \left\{\begin{array}{c}
    |\frac{3}{2},\frac{3}{2}\rangle=\frac{1}{\sqrt2}|p_x+\mathrm{i} p_y,\uparrow\rangle\\
      |\frac{3}{2},\frac{1}{2}\rangle=\frac{1}{\sqrt6}|p_x+\mathrm{i} p_y,\downarrow\rangle-\sqrt{\frac{2}{3}}|p_z,\uparrow\rangle\\
         |\frac{3}{2},-\frac{1}{2}\rangle=-\frac{1}{\sqrt6}|p_x-\mathrm{i} p_y,\uparrow\rangle-\sqrt{\frac{2}{3}}|p_z,\downarrow\rangle\\
            |\frac{3}{2},-\frac{3}{2}\rangle=\frac{1}{\sqrt2}|p_x-\mathrm{i} p_y,\downarrow\rangle\\
               |\frac{1}{2},\frac{1}{2}\rangle=\frac{1}{\sqrt3}|p_x+\mathrm{i} p_y,\downarrow\rangle+\frac{1}{\sqrt 3}|p_z,\uparrow\rangle\\
                  |\frac{1}{2},-\frac{1}{2}\rangle=-\frac{1}{\sqrt3}|p_x-\mathrm{i} p_y,\uparrow\rangle+\frac{1}{\sqrt 3}|p_z,\downarrow\rangle
    \end{array}\right.,
\end{equation}
from which the eight-band Kane model (to first-order) is found to be \cite{kpbook}:
\begin{equation}\label{app:kane}
    H_{Kane}=\left(
    \begin{array}{cccccccc}
    E_0& -\sqrt{\frac{2}{3}}Pq_z& \frac{P}{\sqrt 2}q_+& \frac{1}{\sqrt 3} P q_z& 0& -\sqrt{\frac{1}{6}}Pq_-& 0& -\sqrt{\frac{1}{3}}Pq_-\\
    -\sqrt{\frac{2}{3}}Pq_z& 0& 0& 0& \sqrt{\frac{1}{6}}Pq_-& 0& 0& 0\\
    \frac{P}{\sqrt 2}q_-& 0& 0& 0& 0& 0& 0& 0\\
    \sqrt{\frac{1}{3}}Pq_z& 0& 0& -\Delta_0& \sqrt{\frac{1}{3}}Pq_-& 0& 0& 0\\
    0& \sqrt{\frac{1}{6}}Pq_+& 0& \sqrt{\frac{1}{3}}Pq_+& E_0& -\sqrt{\frac{2}{3}}Pq_z& \frac{P}{\sqrt 2}q_-& \sqrt{\frac{1}{3}}Pq_z\\
    -\sqrt{\frac{1}{6}}Pq_+& 0& 0& 0& -\sqrt{\frac{2}{3}}Pq_z& 0& 0& 0\\
    0& 0& 0& 0& -\sqrt{\frac{2}{3}}Pq_z& 0& 0& 0\\
    -\sqrt{\frac{1}{3}}Pq_+& 0& 0& 0& \sqrt{\frac{1}{3}}Pq_z& 0& 0& -\Delta_0
    \end{array}
    \right).
\end{equation}
 Note that Eq. \ref{app:kane} is obtained in the following basis:
\[|s\uparrow\rangle,|\frac{3}{2},\frac{1}{2}\rangle,|\frac{3}{2},\frac{3}{2}\rangle,|\frac{1}{2},\frac{1}{2}\rangle,|s\downarrow\rangle,|\frac{3}{2},-\frac{1}{2}\rangle,|\frac{3}{2},-\frac{3}{2}\rangle,|\frac{1}{2},-\frac{1}{2}\rangle,\]
of which,
$\{|\frac{3}{2},\frac{3}{2}\rangle,|\frac{3}{2},\frac{1}{2}\rangle,|\frac{3}{2},-\frac{1}{2}\rangle,|\frac{3}{2},-\frac{3}{2}\rangle\}$ belongs to the co-irrep $\{3\}$ of $\Gamma$, $\{|\frac{1}{2},\frac{1}{2}>\rangle,|\frac{1}{2},-\frac{1}{2}>\rangle\}$ belongs to the co-irrep $\{1\}$ of $\Gamma$, while $\{|s\uparrow>\rangle,|s\downarrow \rangle\}$ belongs to the co-irrep $\{2\}$ of $\Gamma$, all from a unitary transformation of our basis vectors to derive the explicit co-irrep matrices we show in SM: Part I \cite{SM}.
This makes our obtained model $H_{216}$ in Eq. \ref{examples-2} different from $H_{Kane}$ in the form. Through the standard method of projection operator in group theory, we find the transformation matrix from basis vectors  deriving Eq. \ref{examples-2} to those deriving Eq. \ref{app:kane}  shown above  is as follows:
\begin{equation}\label{app:Ukane}
U_{Kane}=\left(\begin{array}{cccccccc}
 0 & 0 & 0 & 0 & 0 & 0 & 0 & -1 \\
 0 & 0 & 0 & 1 & 0 & 0 & 0 & 0 \\
 0 & 0 & 0 & 0 & 1 & 0 & 0 & 0 \\
 1 & 0 & 0 & 0 & 0 & 0 & 0 & 0 \\
 0 & 0 & 0 & 0 & 0 & 1 & 0 & 0 \\
 0 & 0 & 1 & 0 & 0 & 0 & 0 & 0 \\
 0 & -1 & 0 & 0 & 0 & 0 & 0 & 0 \\
 0 & 0 & 0 & 0 & 0 & 0 & 1 & 0 \\
\end{array}
\right).
\end{equation}
Then $U_{Kane}^\dag \mathcal{H}_{216} U_{Kane}$ would result in:
\begin{equation}
\mathcal{H}_{216}=\left(
\begin{array}{cccccccc}
 \epsilon_2 & r_4 q_z & -\frac{1}{2} \sqrt{3} r_4 q_+ & -r_2 q_z & 0 & \frac{r_4 q_-}{2 } & 0 &  r_2 q_- \\
 r_4 q_z & \epsilon_3 & - r_1 q_- & 0 & -\frac{r_4 q_-}{2 } &  \sqrt{3} r_1 q_+ &  2 r_1 q_z & \frac{1}{2} \sqrt{3} r_3 q_+ \\
 -\frac{1}{2} \sqrt{3} r_4 q_- & -r_1 q_+ & \epsilon_3 & \frac{r_3 q_+}{2} & 0 & 2 r_1 q_z & \sqrt{3} r_1 q_- & -r_3 q_z \\
 -r_2 q_z & 0 & \frac{r_3 q_-}{2} & \epsilon_1 & -r_2 q_- & \frac{1}{2} \sqrt{3} r_3 q_+ & -r_3 q_z & 0 \\
 0 & -\frac{r_4 q_+}{2 } & 0 & -r_2 q_+ & \epsilon_2 & r_4 q_z & -\frac{1}{2} \sqrt{3} r_4 q_- & -r_2 q_z \\
 \frac{r_4 q_+}{2} &  \sqrt{3} r_1 q_- & 2r_1 q_z & \frac{1}{2} \sqrt{3} r_3 q_- &  r_4 q_z & \epsilon_3 &  r_1 q_+ & 0 \\
 0 & 2r_1 q_z & \sqrt{3} r_1 q_+ & -r_3 q_z & -\frac{1}{2} \sqrt{3} r_4 q_+ & r_1 q_- & \epsilon_3 & -\frac{r_3 q_-}{2} \\
 r_2 q_+ & \frac{1}{2} \sqrt{3} r_3 q_- & -r_3 q_z & 0 & -r_2 q_z & 0 & -\frac{r_3 q_+}{2} & \epsilon_1 \\
\end{array}
\right).\end{equation}
When $\epsilon_3\rightarrow 0, \epsilon_2\rightarrow E_0, \epsilon_1\rightarrow-\Delta_0, r_4\rightarrow -\sqrt{\frac{2}{3}}P, r_2\rightarrow-\frac{1}{\sqrt 3}P$ and $r_1,r_3\rightarrow0$,  $\mathcal{H}_{216}$ would become $H_{Kane}$.

\subsection{Hopf link nodal loops in SG 62}\label{app:detail-62}

We  then show a $k\cdot p$ Hamiltonian which could realize a Hopf-link nodal structure, namely, two nodal loops are nested with each other as shown in Fig. \ref{hopf}. As far as we know, the $k\cdot p$ Hamiltonians in the study of topological semimetals were constructed to show nodal point or one single nodal loop very near to the central $k$ vector: $\mathbf{k}^*$. It seems difficult to construct a $k\cdot p$ model which could realize several nodal loops in relatively large range of the BZ. However,  the theory of full-zone $k\cdot p$ model \cite{kpbook} has  proven feasible. Here we take space group 62 with TRS (MSG 62.442) as example to construct an eight-band model to show the Hopf-link nodal structures \cite{hopf-link-1, hopf-link-2,hopf-link-3}.
The $k\cdot p$ model is constructed around some $k$ point in the high-symmetry line $Q$ connecting high-symmetry points $S$ and $R$ and  double-valued representations are considered. In fact, all the $k\cdot p$ models around the points in $Q$ would share the same form. From SM: Pat I \cite{SM}, we know that $Q$ owns four different two-dimensional co-irreps, denoted as $\{1,1\},\{2,2\},\{3,3\},\{4,4\}$, namely, the 4 irreps of $H(\mathbf{k}^*)$ are all paired with themselves by $A$ and belong to  case 1 in Eq. \ref{app:herring}.

 \subsubsection{Compatibility relations}
Firstly we show the essential existence of hourglass band crossing \cite{hourglass-nature} in high-symmetry line $Q$ from compatibility relations in SG 62 (both TRS and SOC are considered) as shown in the following:
\begin{equation}\label{app:crs}
   \left\{\begin{array}{c}
   S\{1,1\}\rightarrow Q\{2,2\}+Q\{4,4\}\\
   S\{2,2\}\rightarrow Q\{1,1\}+Q\{3,3\}\\
   R\{1,1\}\rightarrow Q\{1,1\}+Q\{2,2\}\\
   R\{2,2\}\rightarrow Q\{3,3\}+Q\{4,4\}\\
   Q\{1,1\}\rightarrow P5\{2,2\}\\
   Q\{2,2\}\rightarrow P5\{1,1\}\\
   Q\{3,3\}\rightarrow P5\{2,2\}\\
   Q\{4,4\}\rightarrow P5\{1,1\}\\
   Q\{1,1\}\rightarrow P6\{1,1\}\\
   Q\{2,2\}\rightarrow P6\{2,2\}\\
   Q\{3,3\}\rightarrow P6\{2,2\}\\
   Q\{4,4\}\rightarrow P6\{1,1\}\\
   \end{array}\right.
\end{equation}
where $Q\{1,1\}$ means the co-irrep $\{1,1\}$ for the high-symmetry line $Q$ and so on.  Here the coordinates of related $k$ points are: $Q(-\frac{1}{2},\frac{1}{2},w)$,  $S(-\frac{1}{2},\frac{1}{2},0)$,   $R(-\frac{1}{2},\frac{1}{2},\frac{1}{2})$,   $P5(u,\frac{1}{2},v)$ ($yz$ plane) and  $P6(-\frac{1}{2},u,v)$ ($zx$ plane). From $S$ to $R$, there must be an hourglass band connectivity  where the necking of the hourglass can contain the co-irreps $Q\{1,1\}$ and $Q\{4,4\}$ or $Q\{2,2\}$ and $Q\{3,3\}$.  No matter the band crossing in $Q$ is of hourglass-type or not, there would be 6 possibilities of co-irrep pairs to constitute a band crossing in $Q$. From concrete pairs of two different co-irreps in $Q$, the resulting band crossing may lie in a nodal loop which can be diagnosed from the compatibilities in Eq. \ref{app:crs}. For example, when the band crossing in $Q$ contains $Q\{2,2\}$ and $Q\{4,4\}$, since they  preserve to be different co-irreps in $P6$, thus the band crossing lies in a nodal loop in $P6$, while they become the same co-irrep of $P5$, so it is hard for a nodal loop in $P5$ to go through this band crossing. The 6 possibilities are listed below:
\[
\begin{array}{lc}
\text{band crossing in } $Q$& \makecell{\text{the position of nodal loop}\\\text{from the band crossing}}\\
$Q\{1,1\}+Q\{2,2\}$:& $P5,P6$\\
$Q\{1,1\}+Q\{3,3\}$:& $P6$\\
$Q\{1,1\}+Q\{4,4\}$:& $P5$\\
$Q\{2,2\}+Q\{3,3\}$:& $P5$\\
$Q\{2,2\}+Q\{4,4\}$:& $P6$\\
$Q\{3,3\}+Q\{4,4\}$:& $P5,P6$\\
\end{array}.
\]

\subsubsection{Construction of the $k\cdot p$ model}
Consider some $k$ point in $Q$, $\mathbf{k}^*$, and consider four energy levels at this point denoted as $E_a=\epsilon_1,E_b=\epsilon_2,E_c=\epsilon_3,E_d=\epsilon_4$ whose co-irreps are $\xi_a=\{1,1\},\xi_b=\{2,2\},\xi_c=\{3,3\},\xi_d=\{4,4\}$,respectively. The zeroth order $k\cdot p$ Hamiltonian is diagonal as $\mathrm{dia}(\epsilon_1,\epsilon_1,\epsilon_2,\epsilon_2,\epsilon_3,\epsilon_3,\epsilon_3,\epsilon_3)$. In the following, we firstly construct the first order $k\cdot p$ Hamiltonian $\mathcal{H}_{L=1}$ considering these eight bands. Let's begin as before one by one. For $\mathcal{H}^1_{aa}$, we should consider $H^1_{1,1,m}$ and $\bar{H}^1_{1,1,\bar{m}}$ based on Table \ref{AA}. From  SM: Part III \cite{SM}, we know that $\bar{H}^1_{1,1,\bar{m}}=\{\}$ while
$H^1_{1,1,m}=\{E_1P_3\}$. Thus, $\mathcal{H}^1_{aa}=\left(\begin{array}{cc} W_{11}=r_1E_1P_3&0\\0& W_{22}\end{array} \right)$, where $W_{22}$ should be $u_1 W_{11}(-A_0^{-1}\mathbf{q})^* {u_{1}}^\dag$. As shown in  SM: Part I \cite{SM}, we know that, $A_0=C_{2y}$ and $u_1=(1),u'_1=(-1)$ (also $u_2=(1),u'_2=(-1)$,$u_3=(1),u'_3=(-1)$,$u_4=(1),u'_4=(-1)$), thus $W_{22}=r_1E_1P_3$. Hence, $\mathcal{H}^1(\mathbf{q})_{aa}=(r_1q_z)I_2$.  Similarly,  $\mathcal{H}^1_{bb}=(r_2q_z)I_2$, $\mathcal{H}^1_{cc}=(r_3q_z)I_2$,$\mathcal{H}^1_{dd}=(r_4q_z)I_2$. Then consider $\mathcal{H}^1_{ab}$. We then should know $\{H^1_{1,2,m}\}$, which is $\{\}$, so that $\mathcal{H}^1_{ab}=0$. Similarly, $\mathcal{H}^1_{cd}=0$. For $\mathcal{H}^1_{ac}$, we should know $\{H^1_{1,3,m}\}$ which is found to be $\{M_{\{1,1\}P_2}\}$, thus $\mathcal{H}^1_{ac}=\left(\begin{array}{cc} c_1q_y& c_2^*q_y\\c_2q_y& -c_1^* q_y  \end{array}\right)$. Similarly, $\mathcal{H}^1_{ad}=\left(\begin{array}{cc} c_3q_x& -c_4^*q_x\\c_4q_x& c_3^* q_x  \end{array}\right)$,
$\mathcal{H}^1_{bc}=\left(\begin{array}{cc} c_5q_x& -c^*_6q_x\\c_6q_x& c_5^* q_x  \end{array}\right)$ and $\mathcal{H}^1_{bd}=\left(\begin{array}{cc} c_7q_y& c^*_8q_y\\c_8q_y& -c_7^* q_y  \end{array}\right)$. Thus the first-order $k\cdot p$ Hamiltonian would be:

\begin{equation}\label{examples-3-1}
\left(
\begin{array}{cccc}
r_1q_zI_2& 0& q_y J_{+}(c_1,c_2)& q_x J_-(c_3,c_4)\\
0& r_2q_zI_2&q_x J_{-}(c_5,c_6)& q_y J_+(c_7,c_8)\\
h.c.& h.c.& r_3q_zI_2& 0\\
h.c.& h.c.& 0& r_4q_zI_2
\end{array}
\right).
\end{equation}
where $J_{\pm}(c,c')=\left(\begin{array}{cc}c& \pm {c'}^*\\c'& \mp c^*\end{array}\right).$

Next consider $L=2$. First, the contribution from $l=0$ is $q^2\mathrm{dia}(r'_9,r'_9,r'_{10},r'_{10},r'_{11},r'_{11},r'_{12},r'_{12})$.   For $\mathcal{H}^2_{aa}$, we should know $\{H^2_{1,1,m}\}$ and $\{\bar{H}^2_{1,1,\bar{m}}\}$, and from  SM: Part III \cite{SM}, we find that the latter is $\{\}$ while the former is, $\{D_1E_1,D_5E_1\}$. Thus $\mathcal{H}^2_{aa}=r'_1\frac{1}{2}(q_x^2-q_y^2)-\frac{r'_2 \left(q_x^2+q_y^2-2 q_z^2\right)}{2 \sqrt{3}}$ (multiplied by $I_2$). Similarly,  $\mathcal{H}^2_{bb}=r'_3\frac{1}{2}(q_x^2-q_y^2)-\frac{r'_4 \left(q_x^2+q_y^2-2 q_z^2\right)}{2 \sqrt{3}}$, $\mathcal{H}^2_{cc}=r'_5\frac{1}{2}(q_x^2-q_y^2)-\frac{r'_6 \left(q_x^2+q_y^2-2 q_z^2\right)}{2 \sqrt{3}}$ and $\mathcal{H}^2_{dd}=r'_7\frac{1}{2}(q_x^2-q_y^2)-\frac{r'_8 \left(q_x^2+q_y^2-2 q_z^2\right)}{2 \sqrt{3}}$. Then consider $\mathcal{H}^2_{ab}$, for which we should know $\{H^2_{1,2,m}\}$ which is $\{D_2M_{\{1,1\}}\}$ thus $\mathcal{H}^2_{ab}=q_xq_y J_+(\lambda_1,\lambda_2)$. Similarly, we can obtain $\mathcal{H}^l_{ac}$,$\mathcal{H}^l_{ad}$,$\mathcal{H}^l_{bc}$,$\mathcal{H}^l_{bd}$ and $\mathcal{H}^l_{cd}$ thus own another 10 complex parameters $\{\lambda_i\}_{i=3}^{12}$.   The explicit form of $2$-order $k\cdot p$ Hamiltonian is as follows,

\begin{equation}\label{examles-3}
    \left(
\begin{array}{cccc}
m_1^\mu q_\mu^2& q_xq_y J_+(\lambda_1,\lambda_2)& q_yq_zJ_+(\lambda_3,\lambda_4)&  q_xq_z J_-(\lambda_5,\lambda_6)\\
h.c.&  m_2^\mu q_\mu^2& q_xq_z J_-(\lambda_7,\lambda_8)&  q_yq_z J_+(\lambda_{9},\lambda_{10})\\
h.c.& h.c.& m_3^\mu q_\mu^2& q_xq_y J_+(\lambda_{11},\lambda_{12})\\
h.c.&h.c.&h.c.&m_4^\mu q_\mu^2
\end{array}
\right),
\end{equation}

where $\mu=x,y,z$ and the Einstein summation rule over $\mu$ has been adopted. Furthermore,  $m^x_1=\frac{r'_1-\frac{r'_2}{\sqrt3}}{2}+r'_{9},
m^y_1=\frac{-r'_1-\frac{r'_2}{\sqrt3}}{2}+r'_{9},
m^z_1=\frac{r'_2}{\sqrt 3}+r'_{9}$, $m^x_2=\frac{r'_3-\frac{r'_4}{\sqrt3}}{2}+r'_{10},
m^y_2=\frac{-r'_3-\frac{r'_4}{\sqrt3}}{2}+r'_{10},
m^z_2=\frac{r'_4}{\sqrt 3}+r'_{10}$,  $m^x_3=\frac{r'_5-\frac{r'_6}{\sqrt3}}{2}+r'_{11},
m^y_3=\frac{-r'_5-\frac{r'_6}{\sqrt3}}{2}+r'_{11},
m^z_3=\frac{r'_6}{\sqrt 3}+r'_{11}$, and  $m^x_4=\frac{r'_7-\frac{r'_8}{\sqrt3}}{2}+r'_{12},
m^y_4=\frac{-r'_7-\frac{r'_8}{\sqrt3}}{2}+r'_{12},
m^z_4=\frac{r'_8}{\sqrt 3}+r'_{12}$.

\begin{table}
\centering
\begin{tabular}{|c|c|c|c|c|c|c|c|}
  \hline
  $\epsilon_1$ & $\epsilon_2$ & $\epsilon_3$ & $\epsilon_4$ & $m^1_x$ & $m^1_y$ & $m^1_z$ & $m^2_x$ \\\hline
 $-1$ & $0$ & $0$ & $1$ & $0.5$ & $0.5$ & $0.5$ & $0$ \\\hline
 $m^2_y$ & $m^2_z$ & $m^3_x$ & $m^3_y$&$m^3_z$ & $m^4_x$ & $m^4_y$ & $m^4_z$ \\\hline
    $0$ & $0$ & $0$ & $0$ & $0$ & $0.5$ & $0.5$ & $0.5$ \\\hline
      $r_1$ & $r_2$ & $r_3$ & $r_4$ & $c_1$ & $c_2$ & $c_3$ & $c_4$ \\\hline
        $0$ & $-0.8$ & $0.8$ & $0$ & $0.3$ & $0.3$ & $0.3$ & $0.3$ \\\hline
        $c_5$ & $c_6$ & $c_7$ & $c_8$ & $\lambda_1$ & $\lambda_2$ & $\lambda_3$ & $\lambda_4$ \\\hline
         $0.3$ & $0.3$ & $0.3$ & $0.3$ & $0$ & $0$ & $0$ & $0$ \\\hline
         $\lambda_5$ & $\lambda_6$ & $\lambda_7$ & $\lambda_8$ &$\lambda_{9}$ & $\lambda_{10}$ & $\lambda_{11}$ & $\lambda_{12}$  \\\hline
         $0$ & $0$ & $0$ & $0$ &$0$ & $0$ & $0$ & $0$  \\\hline
\end{tabular}\caption{Parameters in the Hopf-link model.}\label{para}
\end{table}

The $k\cdot p$ model up to the second order, $\mathcal{H}_{62}$ is thus the sum of Eqs. \ref{examples-3-1} and \ref{examles-3} and the zeroth term:

\begin{equation}
\mathcal{H}_{62}(\mathbf{q})=\left(\begin{array}{cccc}
\epsilon_1+m_1^\mu q_\mu^2+r_1q_z& q_x q_y J_{+}(\lambda_1,\lambda_2) & q_y J_+(c_1,c_2)+q_y q_z J_{+}(\lambda_3,\lambda_4)& q_x J_-(c_3,c_4)+q_x q_z J_-(\lambda_5,\lambda_6)\\
h.c. & \epsilon_2+m_2^\mu q_\mu^2+r_2q_z& q_x J_-(c_5,c_6)+q_x q_z J_-(\lambda_7,\lambda_8)& q_y J_+(c_7,c_8)+q_y q_z J_+(\lambda_9,\lambda_{10})\\
 h.c.& h.c.& \epsilon_3+m_3^\mu q_\mu^2+r_3q_z& q_x q_y J_+(\lambda_{11},\lambda_{12})\\
 h.c.&h.c. &h.c. & \epsilon_4+m_4^\mu q_\mu^2+r_4q_z\\
\end{array}
\right),
\end{equation}\label{62}
where  Einstein summation rule is adopted over $\mu=x,y,z$.

Setting $q_x=q_y=0$, we obtain that $\mathcal{H}_{62}=\bigoplus_{i=1}^4(\epsilon_i+r_i q_z+m_i^z q_z^2)\left(\begin{array}{cc}1& 0\\0& 1\end{array}\right)$ as expected since $\mathbf{q}=(0,0,q_z)$ lies in $Q$.   We further assume that $\mathbf{k}^*$ is chosen to be the hourglass band crossing from bands with co-irreps $\{2,2\}$ and $\{3,3\}$. Thus $\epsilon_2=\epsilon_3$ and they can be  set to be zero. We demonstrate the nested nodal loops shown in Fig. \ref{hopf} after choosing a set of parameters as shown in Table \ref{para}. As seen from Fig. \ref{hopf},  we can find the nodal loops originated from the band crossings in $Q$ are consistent with the predictions from compatibility relations described above. This is understandable since the $k\cdot p$ model is constructed using symmetry analysis by the little group.  We expect  $\mathcal{H}(\mathbf{q})$ in Eq. \ref{examles-3} could provide a reliable  model obtained based on concetre SG  and be applied to further study  interesting physical properties arising from these nodal loops \cite{hopf-link-1,hopf-link-2,hopf-link-3}.
\begin{figure*}[!htbp]
\centering
\includegraphics[width=16 cm]{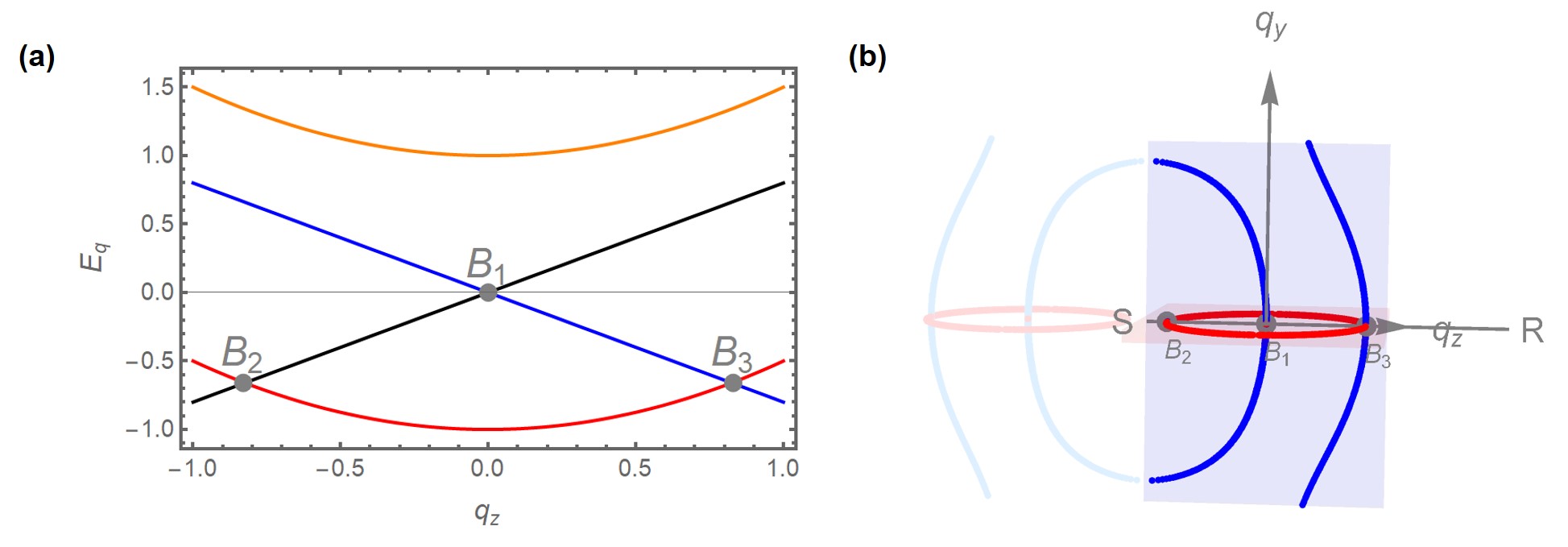}
    \caption{(a) The band structure of $\mathcal{H}(\mathbf{q})$ in Eq. \ref{examles-3} along $q_z$ direction with $q_x=q_y=0$. The red, blue, black, orange colors denote the co-irreps $\{1,1\}$, $\{2,2\}$,$\{3,3\}$, and $\{4,4\}$ respectively. Three band crossings are labeled. $B_1$ is the hourglass band crossing containing co-irreps $\{2,2\}$ and $\{3,3\}$, which can only lie in a nodal loop in $yz$ plane. $B_2$ composes of co-irreps $\{1,1\}$ and $\{3,3\}$ which can only lie in a nodal loop in $zx$ plane. $B_3$ composes of co-irreps $\{1,1,\}$ and $\{2,2\}$, which lie in two nodal loops in $yz$ and $zx$ planes, respectively. (b) The demonstration of several nodal loops. $q_x$ axis is not shown which is perpendicular to the inside of the paper. The high-symmetry line $Q$ connects the high-symmetry points $S$ and $R$. $B_3$ links two nodal loops (in red and blue) while the red nodal loop is nested with the blue nodal loops originated from the band crossing $B_1$. Note that here the  blue nodal loop originated from $B_1$ and  its symmetry (such as time-reversal) related counterpart form  a whole nodal loop. The light blue and light red nodal loops are obtained by inversion around $S$ point.   }\label{hopf}
\end{figure*}

\end{appendix}

\end{widetext}

\end{document}